\documentclass[a4paper,fleqn,usenatbib]{mnras}

\usepackage[T1]{fontenc}
\usepackage{ae,aecompl}

\usepackage{graphicx}	
\usepackage{amsmath}	
\usepackage{amssymb}	

\title[Emission-rotation correlation in pulsars]{Emission-rotation correlation in pulsars: new discoveries with
  optimal techniques}

\author[P. R. Brook et al.]{
P. R. Brook,$^{1,2}$\thanks{E-mail: paul.brook@astro.ox.ac.uk}
A. Karastergiou,$^{1,3}$
S. Johnston,$^{2}$
M. Kerr,$^{2}$
R.M. Shannon$^{2}$
\newauthor
and S. J. Roberts$^{4}$
\\
$^{1}$Astrophysics, University of Oxford, Denys Wilkinson
Building, Keble Road, Oxford, OX1 3RH, UK\\
$^{2}$CSIRO Astronomy and Space Science, Australia
Telescope National Facility, P.O. Box 76, Epping, NSW 1710,
Australia\\
$^{3}$Physics Department, University of the Western Cape, Cape Town
7535, South Africa\\
$^{4}$Information Engineering, University of Oxford, Parks Road, Oxford, OX1 3PJ, UK
}

\date{Accepted XXX. Received YYY; in original form ZZZ}

\pubyear{2015}

\begin{document}
\label{firstpage}
\pagerange{\pageref{firstpage}--\pageref{lastpage}}
\maketitle

\begin{abstract}
Pulsars are known to display short-term variability. Recently,
examples of longer-term emission variability have emerged that are often correlated with
changes in the rotational properties of the pulsar. To further
illuminate this relationship, we have developed techniques to identify
emission and rotation variability in pulsar data, and determine
correlations between the two. Individual observations may be too noisy
to identify subtle changes in the pulse profile. We
use Gaussian process (GP) regression to model noisy observations and
produce a continuous map of pulse profile variability. Generally,
multiple observing epochs are required to obtain the pulsar spin
frequency derivative ($\dot{\nu}$). GP regression is,
therefore, also used to obtain $\dot{\nu}$, under the hypothesis that
pulsar timing noise is primarily caused by unmodelled changes in
$\dot{\nu}$. Our techniques distinguish between two types of
variability: changes in the total flux density versus changes in the
pulse shape. We have applied these techniques to 168 pulsars observed
by the Parkes radio telescope, and see that although variations in
flux density are ubiquitous, substantial changes in the shape of the
pulse profile are rare. We reproduce previously published results and present examples of profile
shape changing in seven pulsars; in particular, a clear new example of correlated changes in profile shape and
rotation is found in PSR~J1602$-$5100.  In the shape changing pulsars, a more
complex picture than the previously proposed two state model
emerges. We conclude that our simple assumption that all timing noise can
be interpreted as $\dot{\nu}$ variability is insufficient to explain
our dataset.
\end{abstract}




\section{Introduction}
Pulsars are employed as precision timing tools, due to the stability
of their emission and of their rotation. The average radio pulse shape
(known as the pulse profile) has traditionally been thought to remain
steady over decades, and extreme rotational stability results from the
high angular momentum of a rapidly spinning, dense star. Pulsar
stability exists even in the presence of nulling and mode-changing:
short-term emission variations that were first observed soon after the
discovery of pulsars
\citep{1970Natur.228...42B,1970Natur.228.1297B}. These discontinuous
changes occur on time-scales ranging from a few pulse periods to hours
and days \citep{2007MNRAS.377.1383W}. Mode-changing pulsars switch
between two or more quasi-stable states, while nulling is thought to
be an extreme form of mode-changing, with one state showing no or low
emission. These effects aside, the pulse profile has long been
considered a stable characteristic of each pulsar. The last decade,
however, has witnessed a number of counter-examples to this perceived
stability.
\\
A small, emerging population of radio pulsars have shown pulse profile
changes on time-scales of months to years
\citep[e.g.][]{2010Sci...329..408L}. These changes are often
accompanied by \emph{timing noise}, a term given to the unexplained,
systematic deviation from the modelled rotational behaviour of a
pulsar, often seen in younger pulsars. Although common, the causes of timing
noise are poorly understood. One possibility is that timing noise is
due (at least in part) to unmodelled variability in the spin-frequency
derivative or \emph{spindown rate} ($\dot{\nu}$) of the pulsar. Under this
assumption, timing noise can be modelled as a time-variable
$\dot{\nu}$, which leads to the observed correlation between pulse
profile and $\dot{\nu}$ variations
\citep{2010Sci...329..408L,2013MNRAS.432.3080K,2014ApJ...780L..31B}.
\\
An extreme example of correlated emission and rotation variability is
observed in a group of intermittent pulsars
\citep{2006Sci...312..549K,2012ApJ...746...63C,
  2012ApJ...758..141L}. Such objects cycle quasi-periodically between
intervals in which the pulsar is emitting normally, and those in which
no emission is detected. This cyclic behaviour occurs on time-scales of
months to years. All known intermittent pulsars lose rotational energy
at a much higher rate when their emission is visible; one possible
explanation is changing magnetospheric currents
\citep{2006Sci...312..549K}.
\\
Neutron star glitches, characterised by a discrete increase and
gradual relaxation of the rotational frequency, have also recently
been linked to pulse profile variability in radio pulsars
\citep{2011MNRAS.411.1917W, 2013MNRAS.432.3080K}, providing additional
links between emission and rotation. Furthermore, examples of glitches
and irregular spindown properties associated with emission
variability, have been seen in magnetars \citep{2006csxs.book..547W},
where dramatic profile changes, related to changes in the magnetic
field structure have been observed both in X-rays and radio
\citep[e.g.][]{2007ApJ...663..497C}.
\\
Any unmodelled variability is detrimental to experiments that rely on
precision pulsar timing, such as the search for gravitational waves
using pulsar timing arrays. The observed correlation between timing
and pulse profile variability suggests that this information may be
used to improve the precision of pulsar timing experiments. In
addition, it is revealing a new type of phenomenology that may hold
information on the interiors and environments of pulsars.
\\
In this work we make the assumption that all unmodelled timing
variations can be attributed to changes in $\dot{\nu}$. Under this
assumption we investigate the pulse profile and timing variability in
a large number of pulsars to identify any correlation. We have
analysed data from 168 pulsars that have been monitored for up to
eight years by the Parkes radio telescope. These objects represent a
population of young, energetic pulsars, known to show the most timing
noise, thus offering possibilities to test the above hypothesis. The
only exception is PSR~J0738$-$4042, which is monitored due to its known
variable behaviour \citep{2014ApJ...780L..31B}. We present nine
interesting examples of pulsar variability in this paper, including
PSR~J0738$-$4042 and two pulsars previously studied in
\citet{2010Sci...329..408L}.
\\
In Section 2, we detail the observations. Section 3 describes the data
analysis techniques used to detect both emission and rotational
variability. The results from nine pulsars are presented in Section 4,
followed by a discussion and conclusions in sections 5 and 6
respectively.
\section{Observations}
Since 2007, 168 pulsars have been observed on a roughly monthly basis
at 1369 MHz with the Parkes radio telescope and the Multibeam receiver
as part of the Fermi timing programme \citep{2010PASA...27...64W}. The
data were recorded with one of the Parkes Digital Filterbank systems
(PDFB1/2/3/4) with a total bandwidth of 256 MHz in 1024 frequency
channels. Radio frequency interference was removed using
median-filtering in the frequency domain and then manually excising
bad sub-integrations. Flux densities have been calibrated by
comparison to the continuum radio source 3C 218. The data were then
polarization-calibrated for both differential gain and phase, and for
cross coupling of the receiver. The MEM method based on long
observations of 0437-4715 was used to correct for cross coupling
\citep{2004ApJS..152..129V}. Flux calibrations from Hydra A were used
to further correct the bandpass. After this calibration, profiles were
formed of total intensity (Stokes parameter I), and averaged over time
and frequency.  The profiles were cross-correlated with templates with
a high signal-to-noise ratio (S/N) to obtain times of arrival (TOAs), using
standard techniques for pulsar timing \citep{2006MNRAS.369..655H}.
The template used to calculate the TOAs is noiseless, generated iteratively from
a set of von Mises functions in order to represent the profile, formed
from the summation of all observations.
\section{Data Analysis}
\label{dataanalysis}
The objective of the data analysis is to model the variability of
pulse profiles that have been sampled at irregular intervals, and
compute the $\dot{\nu}$ timeseries for each pulsar. We have developed
a technique that models the pulse profile data as a function of time,
allowing us to interpolate between the epochs of observation. This
builds on work by \citet{2014ApJ...780L..31B}. This process also
allows better visualisation of sparsely sampled data. Irregular
sampling again raises difficulties in the calculation of $\dot{\nu}$,
which we have also addressed using similar inference techniques. These
techniques are based on Gaussian process (GP) non-parametric
modelling, details of which are provided in the following.
\subsection{Pulse profile variability maps}
\label{varmaps}
We use the term \emph{variability map} to describe an interpolated
plot that maps the differences (i.e. the \emph{profile residuals})
between the pulse profile at each observation and a constant
model. The model is taken to be the median of all observed profiles in
a dataset. Panels A and B of Figure~\ref{0738_map} are examples of
variability maps. Before the observed pulse profiles can be compared
with the constant model, we process them to ensure the off-pulse
baseline is centred on zero. Some observations are considered
unreliable and excluded from further analysis. A pulse profile is
excluded using the following empirical rule which has been found to
perform well in rejecting spurious observations: if the standard
deviation of the off-pulse region is greater than a factor of two
larger than the median value taken from the off-pulse regions across
all epochs. Observations are also removed manually if they show
extreme and isolated profile deviations that can likely be attributed
to instrumental failure.
\\
All pulse profiles originally consist of 1024 phase bins across the
pulse period. If S/N is low for any pulsar then pulse profile
variations can become difficult to detect. In cases where the profile
with the highest S/N in a pulsar dataset has a peak value less than
20 times the standard deviation of the off-pulse noise, the S/N is
increased by reducing the number of phase bins to 128. The individual
profiles are aligned by cross-correlation with the median over all
epochs. Using the timing information to align the profiles is not
possible, given the amount of timing noise in the data. Aligning the
profiles is essential for the modelling that follows, as the
timeseries of each pulse phase bin is modelled independently.
In the few observations where large profile deviations occur, it is
possible that the alignment may be slightly biased in that direction,
however, all observations have been individually inspected to exclude
the possibility that this effect plays an important role.
\\
We use variability maps to monitor two types of profile variability:
changes in the flux density across the whole pulse profile, and
changes in the relative flux density of profile components, i.e. shape
variations. In this paper, we primarily focus on the latter, as large
flux density variations are observed in most of the pulsar data
analysed in this work, and are thought to be primarily attributable to
the effects of refractive scintillation. To uncover the less common
pulse profile shape changes, we normalise all observations by the mean
on-pulse flux density.
\\
Two median profiles were calculated for each pulsar dataset: one for
the normalised data described above and one for the non-normalised,
flux calibrated data. The relevant median profile was then subtracted
from each observation, leaving the profile residual \citep[following][]{2014ApJ...780L..31B}. For each of the pulse profile phase
bins, we compute a GP regression model that best describes the profile
residuals \citep{Rasmussen&Williams, Roberts_etal}, which we observe
to often have sharp turnover features.  The covariance
function chosen for this analysis, therefore, employs a kernel from the Matern
class:
\begin{equation}
k(x_{i},x_{j}) = \sigma_{f}^{2}
\frac{2^{1-\mu}}{\Gamma(\mu)}\left(\frac{\sqrt{2\mu}d}{\lambda}\right)^{\mu}K_{\mu}\left(\frac{\sqrt{2\mu}d}{\lambda}\right),
\end{equation}
where $\Gamma$ is the gamma function, $K_{\mu}$ is a modified Bessel function, d is the distance
$|x_{i}-x_{j}|$ between any two epochs (training points),
$\sigma_{f}^{2}$ is the maximum allowable covariance, and $\lambda$ is
the characteristic lengthscale, i.e. a parameter which reflects how
significantly the distance between $x_{i}$ and $x_{j}$ affects
$k(x_{i},x_{j})$. The positive covariance parameter $\mu$ was chosen
to be 3/2 to provide a level of smoothness and flexibility to the
covariance function that is suitable for the kind of trends we are
trying to model \citep{Rasmussen&Williams}:
\begin{equation}
k_{3/2}(x_{i},x_{j}) = \sigma_{f}^{2}\left(1+\frac{\sqrt{3}d}{\lambda}\right)\exp\left(-\frac{\sqrt{3}d}{\lambda}\right).
\end{equation}

This Matern covariance kernel was combined with a white noise kernel
$\sigma_{n}^{2}\delta_{x_{i} x_{j}}$ to model the uncertainty in the
profile data, where $\delta_{x_{i} x_{j}}$ is the Kronecker delta
function and $\sigma_{n}$ is the standard deviation of the noise term
in the model. The choice of hyperparameters $\theta$ ($\sigma_{f}$,
$\lambda$, $\sigma_{n}$) for the covariance function employed by the
GP, is made by maximising log $p(\textbf{y}|\textbf{x},\theta)$. In
this process we constrain the lengthscale $\lambda$ between 30 and 300
days, which we find to result in the models best representing the
data. The GP takes the training points (the observed data; $x_{i}$ and
corresponding $y_{i}$) and calculates test points, i.e. the most
likely value $y_{*}$ for any value $x_{*}$, and its variance:
\begin{equation} \label{ystar}
y_{*} = K_{*}K_{i j}^{-1}\textbf{y}
\end{equation}
\begin{equation} \label{var}
var(y_{*}) = K_{**}-K_{*}K_{ij}^{-1}K_{*}^T,
\end{equation}
where $K_{i j}$ is a covariance matrix with components
$k(x_{i},x_{j})$ over all training points, $K_{*}$ is a matrix which
reflects covariance between a test point and the training points and
has components $k(x_{*},x_{i})$. The covariance of a test point
$K_{**} = k(x_{*},x_{*})$.
\\
A GP regression model is produced for all phase bins, with test points
computed at one day intervals. These models are then combined to
produce a continuous variability map, which highlights deviations
across the pulse profile and across the data span (e.g. Panels A and B
of Figure~\ref{0738_map}).
\subsection{Rotational Variability}
\label{rotvar}
\begin{figure*}
\begin{center}
\includegraphics[width=150mm]{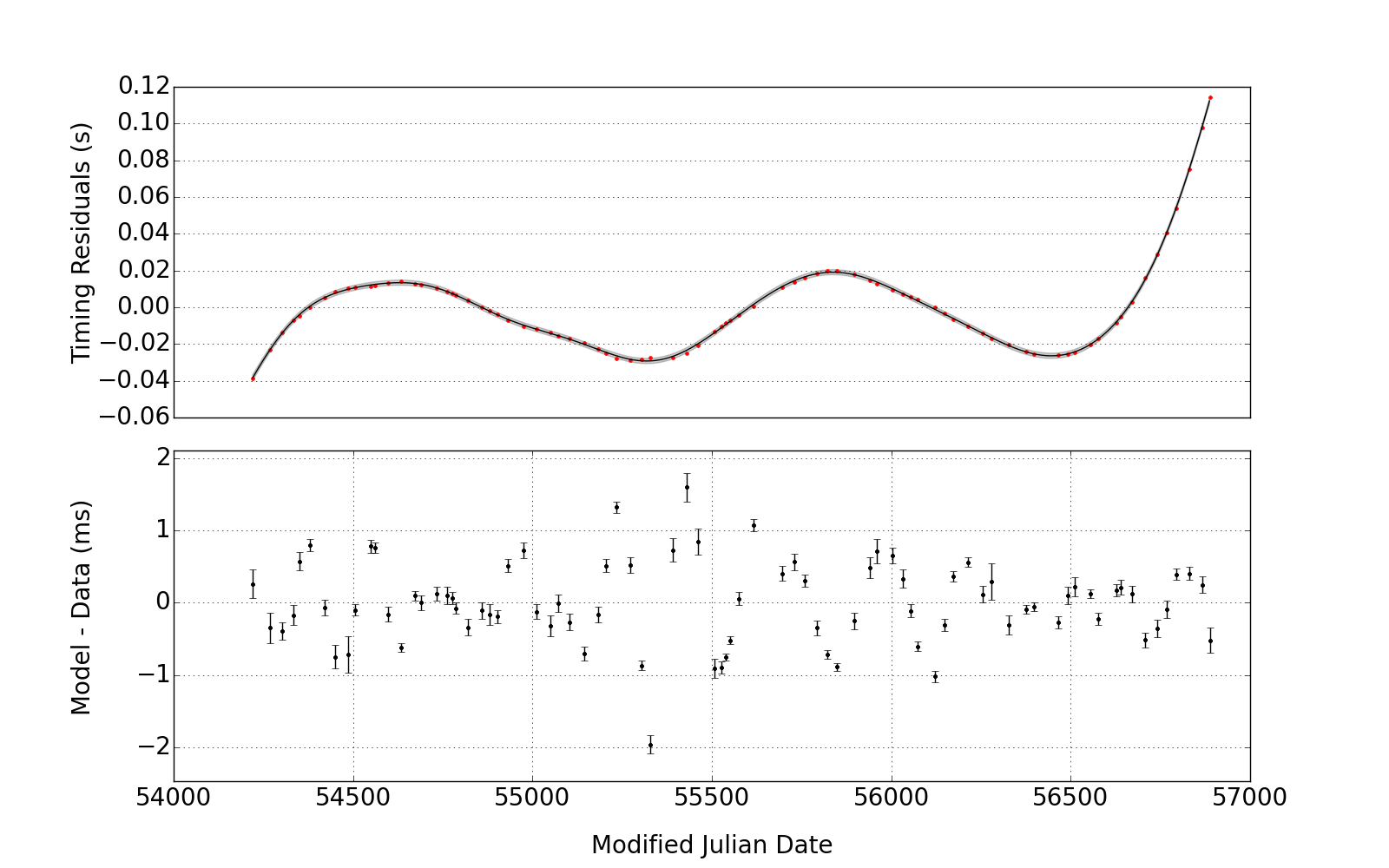}
\caption{The timing residuals and GP model for PSR~J0940$-$5428 using one
  kernel in the covariance function. Top panel: The red points are the
  timing residuals. The black trace shows the GP model, which has a
  covariance function that employs one kernel with a lengthscale of
  276 days. The shaded 2$\sigma$ uncertainty region indicates the range of GP models
that can describe the data. Bottom panel: The GP model minus the timing residuals at the epochs of the observations. The
  structure seen here implies an ill-fitting model that does not
  account for the short-term periodic behaviour seen in the data. The uncertainty in bottom panel is that of
the timing residuals.}
\label{1kern}
\end{center}
\end{figure*}

\begin{figure*}
\begin{center}
\includegraphics[width=150mm]{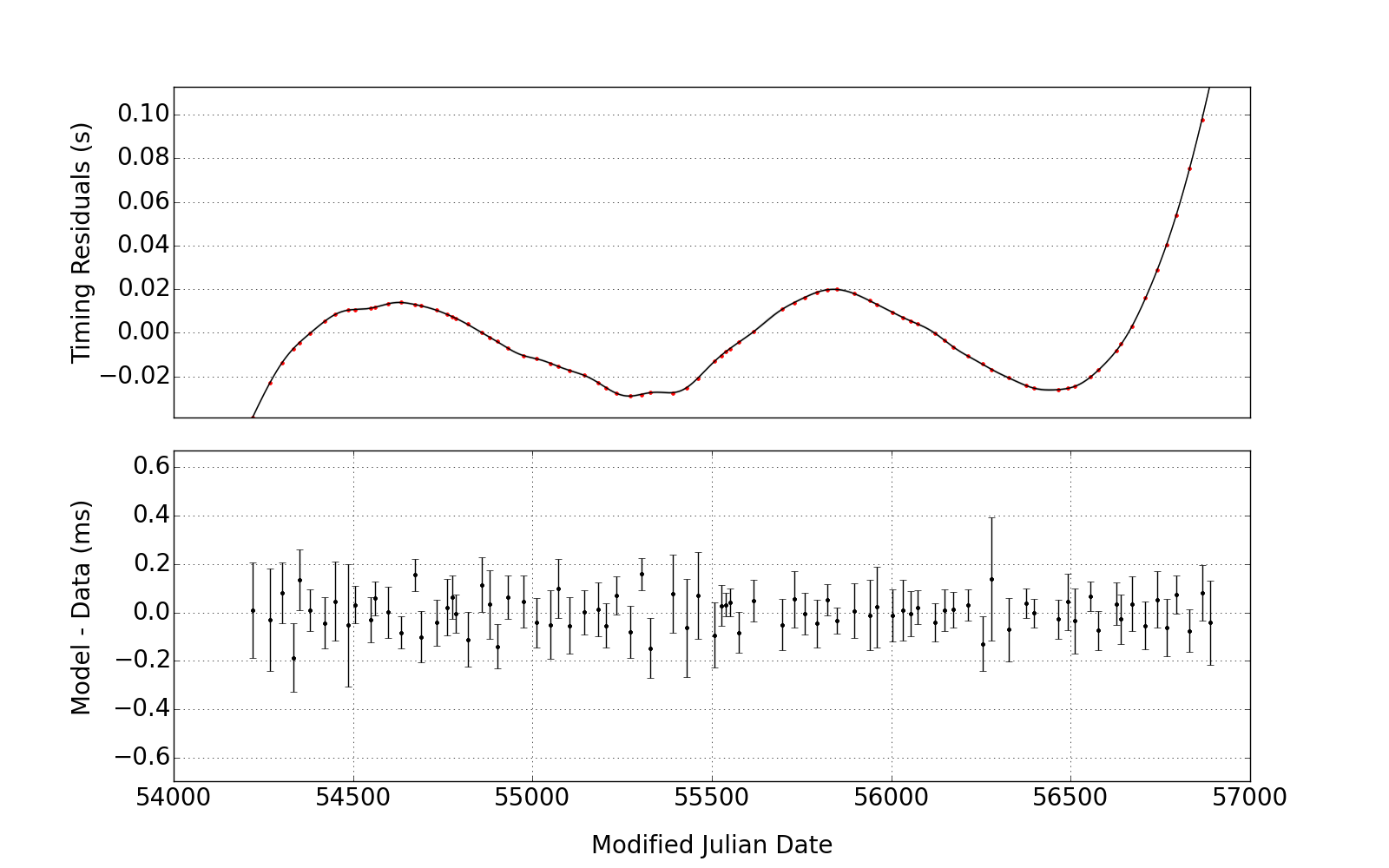}
\caption{The timing residuals and GP model for PSR~J0940$-$5428. As Figure
\ref{1kern}, but employing two kernels in the covariance
function; the lengthscales are 60 days and 471 days. The lack of
structure in the bottom panel suggests a well-fitting model and justifies the number of kernels
and optimised parameters used in the covariance function.}
\label{2kern}
\end{center}
\end{figure*}

As mentioned in the introduction, we are testing the hypothesis that
all unmodelled variations in pulsar timing can be explained as
time-variable changes in $\dot{\nu}$. In order to calculate a time
variable $\dot{\nu}$, we use the \emph{timing residuals}, which are
the differences between observed pulse TOAs and a
timing model with a constant set of parameters (e.g. pulse period,
period derivative, position, proper motion).  The model is optimised
by using a least-squares-fitting procedure to minimize the timing
residuals.
\\
In the case of a pulsar with time variable $\dot{\nu}$, a timing model
with constant parameters will result in systematic timing
residuals. \citet{2013MNRAS.432.3080K} describe how the second
derivative of these timing residuals represents a $\dot{\nu}$ term
that is additional to the timing model. We have developed a new
technique which employs GP regression to analytically model the timing
residuals, allowing us to produce a continuous function representing
$\dot{\nu}$ for each pulsar. The Keith et al. calculation is based on the interpolation
technique of \citet{2012MNRAS.424..244D}, who use a generalised Wiener filter
as a maximum likelihood estimator for pulsar timing residuals. This
technique can be considered a special case of our fully Bayesian
inference method \citep{Rasmussen&Williams}. While
the Deng et al. technique provides a means to smoothly interpolate
between residuals, it does not provide uncertainties for the
interpolated data. This makes it non-trivial to obtain useful
uncertainty estimates of the derivative values of a model produced by
maximum likelihood estimation; Keith et al. find the second derivative
of their timing residual interpolation by numerically differentiating
twice. In contrast, Gaussian process regression allows us to
analytically model the second derivative directly from the timing
residuals, with associated fully Bayesian error estimation. GP
regression does not require regularly evenly sampled data.
\\
For this technique, we use a squared exponential
covariance kernel because it is infinitely differentiable:
\begin{equation}
k(x_{i},x_{j}) =
\sigma_{f}^{2}\exp\left(\frac{-d^{2}}{2\lambda^{2}}\right),
\end{equation}
along with the white noise kernel. When observing how well the GP
model and optimised covariance hyperparameters $\theta$ fit to the
timing residuals, however, we noticed that one kernel was not always
sufficient to describe them. Adding a second squared exponential
kernel to the covariance function models the timing residuals more
accurately in all but two of the nine pulsars detailed in this
work. As an example, Figures \ref{1kern} and \ref{2kern} show the
timing residuals of PSR~J0940$-$5428 fitted with a GP using one and two
kernels respectively. The discrepancies between the data and the model
are also shown. The error bars in the lower panel represent
$\sigma_{TOA}$, the uncertainty in the TOA measurement.  It is clear
that the single kernel model in Figure \ref{1kern} systematically
deviates from the TOA measurements, whereas the same panel of Figure
\ref{2kern} reveals a near perfect fit.
\\
To calculate $\dot{\nu}$, we optimised the hyperparameters $\theta$,
as described in Section \ref{varmaps}. Whether the covariance function
contains one or two kernels, the value of each lengthscale was
restricted to between 30 and 1000 days; all other parameters were
unbounded. The optimised covariance hyperparameters $\theta$ were then
carried forward to calculate the second derivative of the GP
regression model of the timing residuals; following \citet{holsclaw},
the second derivative of a GP model can be estimated using the second
derivative of the covariance kernel. In the case of the squared
exponential kernel:
\begin{equation}
k^{''}(x_{i},x_{j}) = \frac{\sigma_{f}^2}{\lambda^{2}}
\exp\left(\frac{-d^{2}}{2\lambda^{2}}\right)\left(1 - \frac{d^{2}}{\lambda^{2}}\right).
\end{equation}
The second derivative of the GP model:
\begin{equation}
\frac{d^{2}y_{*}}{dx^{2}} = K^{''}_{*}K_{i j}^{-1}\textbf{y}.
\end{equation}
The value of $\dot{\nu}$ can trivially be shown to be:
\begin{equation}
\nu \frac{d^{2}y_{*}}{dx^{2}}.
\end{equation}
\\
The variance of this method is given by:
\begin{equation}
var\left(\frac{d^{2}y_{*}}{dx^{2}}\right) =A - K^{''}_{*}K_{i
  j}^{-1}K^{''}_{*},
\end{equation}
where $A$ is the diagonal terms of $k^{''''}(x_{i},x_{j})$, which gives:
\begin{equation}
\frac{3 \sigma_{f}^2}{\lambda^{4}}
\label{var}
\end{equation}
in the case of one covariance kernel. When plotting the $\dot{\nu}$
model, test points were calculated only for days on which the pulsar
was observed.
\\
\begin{table*}
\begin{center}
\begin{tabular}{cccccccc}
\hline
\hline
PSR~J&PSR B&$\nu$ (s$^{-1}$)&$\dot{\nu}$ (s$^{-2}$)&DM
(cm$^{-3}$pc)& Age (Yr)&$B_{surface}$ (G)&$\dot{E}$ (Js$^{-1})$
\\
\hline
0738$-$4042 &0736$-$40&2.667&$-1.150\times 10^{-14}$&160.8&$3.68\times 10^{6}$&$7.88\times 10^{11}$&$1.2\times 10^{33}$
\\
0742$-$2822&0740$-$28&5.997&$-6.049\times 10^{-13}$&73.782 &
$1.57\times 10^{5}$&$1.69\times 10^{12}$&$1.4\times 10^{35}$
\\
0908$-$4913&0906-49&9.367&$-1.329\times 10^{-12}$&180.37&$1.12\times 10^{5}$&$1.29\times 10^{12}$&$4.9\times 10^{35}$
\\
0940$-$5428&$-$&11.423&$-4.289\times 10^{-12}$&134.5&$4.22\times 10^{4}$&$1.72\times 10^{12}$&$1.9\times 10^{36}$
\\
1105$-$6107&$-$&15.825&$-3.963\times 10^{-12}$&271.01&$6.33\times 10^{4}$&$1.01\times 10^{12}$&$2.5\times 10^{36}$
\\
1359$-$6038&1356$-$60&7.843&$-3.899\times 10^{-13}$&293.71 &
$3.19\times 10^{5}$&$9.10\times 10^{11}$&$1.2\times 10^{35}$
\\
1600$-$5044&1557$-$50&5.192&$-1.365\times 10^{-13}$&260.56 &
$6.03\times 10^{5}$&$9.99\times 10^{11}$& $2.8\times 10^{34}$
\\
1602$-$5100&1558$-$50&1.157&$-9.316\times 10^{-14}$&170.93 &
$1.97\times 10^{5}$&$7.85\times 10^{12}$&$4.3\times 10^{33}$
\\
1830$-$1059&1828$-$11&2.469&$-3.659\times 10^{-13}$&161.50 &
$1.07\times 10^{5}$&$4.99\times 10^{12}$&$3.6\times 10^{34}$
\\
\hline
\hline
\end{tabular}
\end{center}
\caption[]{Properties of the pulsars showing notable variability.}
\end{table*}
\begin{table*}
\begin{center}
\begin{tabular}{ccccccccccc}
\hline
\hline
Pulsar&Max. Cov.&Len.&Noise RMS&Max. Cov. i&Len. i&Max. Cov. ii&Len. ii&Noise
RMS&Mean TOA\\
&1 Kernel&1 Kernel&1 Kernel&2 Kernels&2 Kernels&2 Kernels&2
Kernels&2 Kernels&RMS\\
&(Secs.)&(Days)&(Secs.)&(Secs.)&(Days)&(Secs.)&(Days)&(Secs.)&(Secs.)\\
\hline
\underline{J0738$-$4042}&$4.0\times 10^{-3}$&321.9&$1.3\times 10^{-4}$&$4.0\times 10^{-6}$&185.6&$1.7\times 10^{-2}$&893.0&$1.3\times 10^{-4}$&$2.1\times 10^{-5}$\\
J0742$-$2822&$3.7\times 10^{-2}$&121.4&$2.7\times 10^{-4}$&$3.6\times 10^{-1}$&633.4&$2.2\times 10^{-5}$&59.5&$1.7\times 10^{-4}$&$1.1\times 10^{-5}$\\
J0908$-$4913&$6.1\times 10^{-5}$&125.1&$8.3\times 10^{-5}$&$6.0\times 10^{-7}$&73.0&$1.2\times 10^{-4}$&229.6&$6.5\times 10^{-5}$&$2.4\times 10^{-6}$\\
J0940$-$5428&$6.3\times 10^{-3}$&275.5&$6.5\times 10^{-4}$&$3.1\times 10^{-2}$&471.3&$2.2\times 10^{-6}$&59.5&$1.1\times 10^{-4}$&$1.1\times 10^{-4}$\\
\underline{J1105$-$6107}&$6.1\times 10^{-4}$&120.4&$6.0\times 10^{-4}$&$2.7\times 10^{-3}$&234.3&$2.7\times 10^{-6}$&51.4&$4.7\times 10^{-4}$&$3.1\times 10^{-5}$\\
J1359$-$6038&$8.4\times 10^{-4}$&233.6&$6.3\times 10^{-5}$&$9.5\times 10^{-7}$&124.0&$1.2\times 10^{-2}$&780.9&$4.7\times 10^{-5}$&$5.0\times 10^{-6}$\\
J1600$-$5044&$1.6\times 10^{-4}$&395.2&$1.2\times 10^{-4}$&$5.6\times 10^{-7}$&118.0&$2.4\times 10^{-4}$&615.2&$6.0\times 10^{-5}$&$4.4\times 10^{-6}$\\
J1602$-$5100&$2.2\times 10^{-1}$&166.3&$3.0\times 10^{-4}$&$4.7\times 10^{-5}$&95.8&$5.6\times 10^{+0}$&506.9&$2.4\times 10^{-4}$&$2.2\times 10^{-5}$\\
J1830$-$1059&$1.1\times 10^{-3}$&108.5&$4.8\times 10^{-4}$&$2.1\times 10^{-5}$&61.8&$1.9\times 10^{-3}$&227.9&$3.4\times 10^{-4}$&$3.8\times 10^{-5}$\\
\hline
\hline
\end{tabular}
\end{center}
\caption[]{The optimised parameters (maximum covariance, lengthscale and noise rms) for a covariance function
  with one and two kernels, along with the mean TOA rms for each pulsar. An underlined pulsar name indicates that one
  kernel was used to calculate $\dot{\nu}$. The $\dot{\nu}$ model for
  all other pulsars was produced using two kernels.}
\label{gpparam}
\end{table*}
\subsection{Correlation maps}
\citet{2010Sci...329..408L} show a correlation between $\dot{\nu}$ and
various metrics of profile shape. In order to understand which regions
of the profile may be correlated and whether changes are coincident or
occur with a lag, we calculate Spearman's rank correlation coefficient
(SRCC) between the timeseries of $\dot{\nu}$ and of the profile
residuals in each phase bin of the pulse profile. We choose SRCC, as
it is a measure of the monotonic relationship between the variables,
yet linearity is not required.
\\
We apply a lag between each pair of timeseries, up to 500 days in both
directions and calculate how SRCC changes. The result is a
\emph{correlation map} for each pulsar (e.g. Panel C of
Figure~\ref{0738_profile}). In these maps, we can identify regions of
the pulse profile that show most correlation with $\dot{\nu}$, and
gain information about the temporal relationship between the two
timeseries.
\\
To permit the correlation calculations, we require the two timeseries
to be sampled at the same rate. The GP model of the profile residuals
provides a value at daily intervals. We, therefore, produce
$\dot{\nu}$ models (as described in Section~\ref{rotvar}) which are
also sampled at one day intervals. 
\\
SRCC is only calculated for the pulse phase bins that lie within the
on-pulse region. We have defined this to be the region where the flux
density of the median
pulse profile of a dataset is greater than 1/30 of its peak value.
\section{Results}
We have applied the techniques described in Section~\ref{dataanalysis}
to data from 168 pulsars observed by the Parkes radio telescope. In
the following, we first show results from pulsars with previously
documented variability. We then present new examples, discovered using
our techniques. Finally, we comment on the remainder of the pulsars in
the dataset.
\subsection{Known Variable Pulsars}
\subsubsection{PSR~J0738$-$4042 (B0736$-$40)}
\begin{figure*}
\begin{center}
\includegraphics[width=150mm]{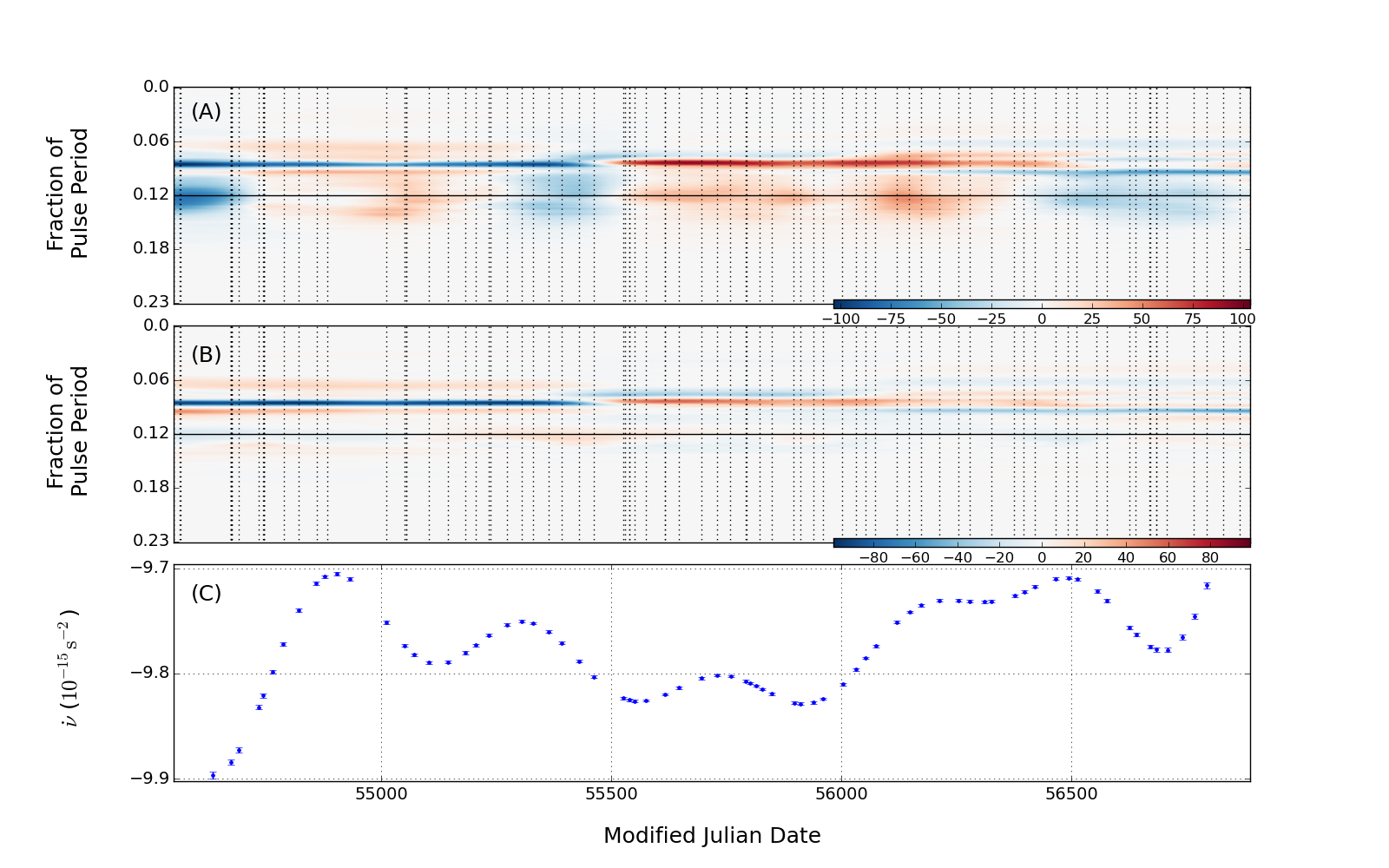}
\caption{Pulse profile and spindown variability for PSR~J0738$-$4042. Panel A: Map showing the difference between the flux
calibrated observations and the median profile across the
dataset. The units are the median of the standard deviation of all
off-peak regions of the dataset. The solid horizontal line highlights the
profile peak, and the vertical dashed lines show the dates of the
included observations. Panel B: As Panel A, but the
observations are first normalised by the
total flux density of the on-pulse region(s). Panel C:
Value of $\dot{\nu}$ on observation dates.}
\label{0738_map}
\end{center}
\end{figure*}

\begin{figure*}
\begin{center}
\includegraphics[width=150mm]{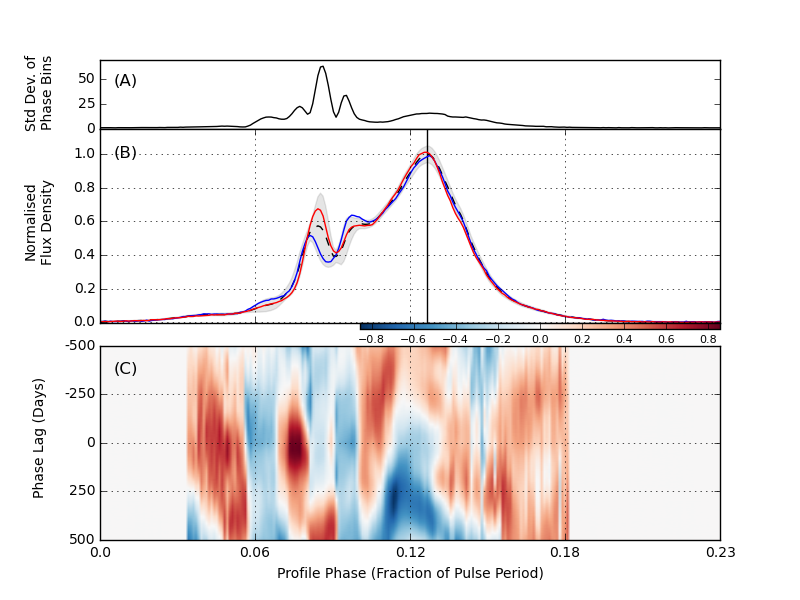}
\caption{Pulse profile variability and correlation map for PSR~J0738$-$4042. 
Panel A: The standard deviation of the data in each profile
  phase bin, in units of the mean standard deviation of the off-pulse
  phase bins. We define an off-pulse phase bin to be one in which the median pulse
  profile of a dataset is less than 1/30 of its peak. Panel B: Black dashed line
  traces the median of normalised pulse profiles across all included
  observations. The blue and red profiles are examples that show the
  extent of shape changes. The blue profile was observed on MJD 54548,
  red on MJD 55616. The shaded area denotes 2$\sigma$ above and below
  the median profile. The solid vertical line marks the peak of the
  profile. Panel C: SRCC for $\dot{\nu}$ and pulse profile
  variability as a function of the pulse phase and lag between the two
  timeseries. A negative lag means that $\dot{\nu}$ is lagging the
  flux density variability.}
\label{0738_profile}
\end{center}
\end{figure*}

\begin{figure*}
\begin{center}
\includegraphics[width=140mm]{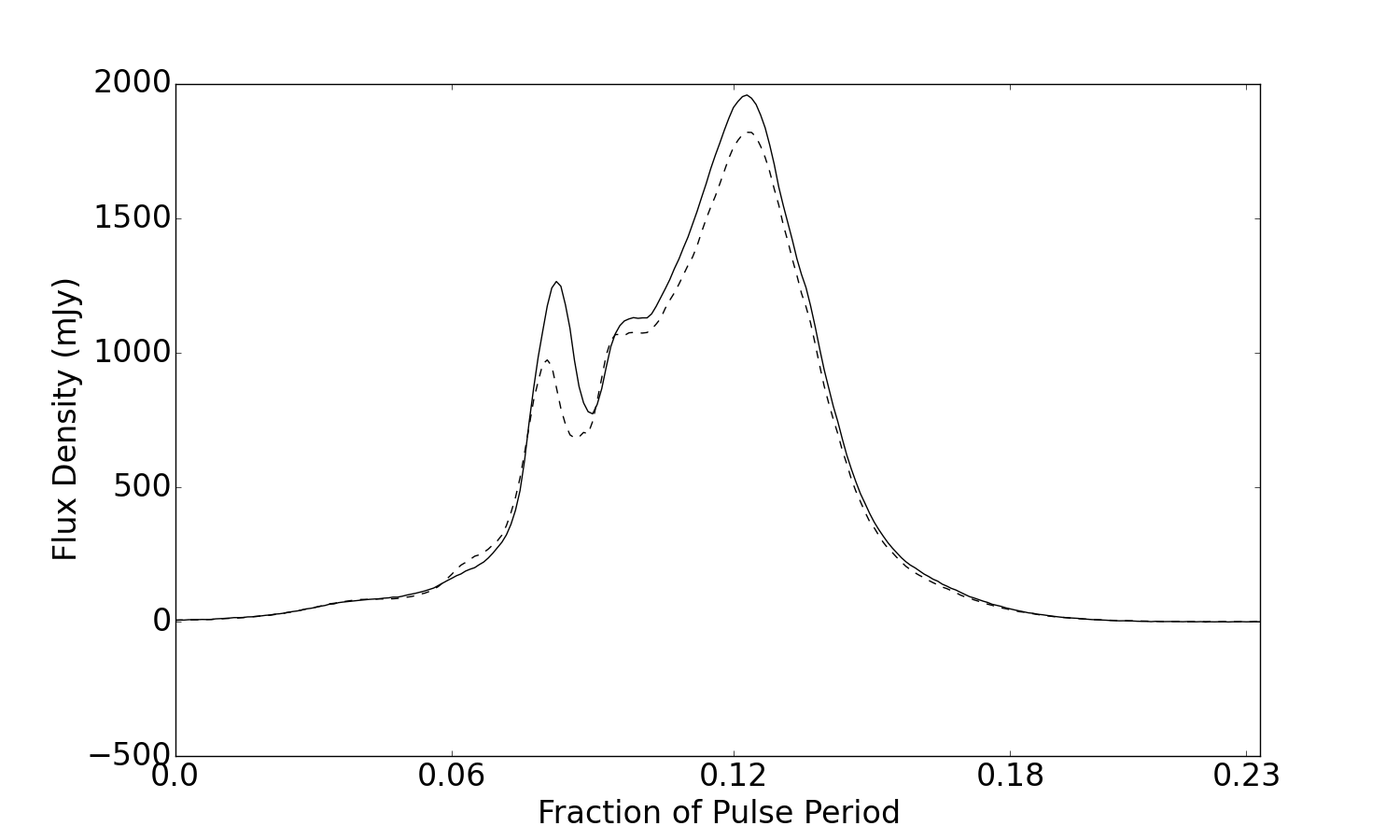}
\caption{Links between the profile shape and the flux density of PSR~J0738$-$4042. The
  solid line traces the median of pulse profiles that were observed
  between MJD 55525 and MJD 56492, i.e. the epochs over which the
  transient leading edge component is at its most prominent relative to the
  profile peak. The
  dashed line traces the median of the pulse profiles that fall
  outside this epoch. The solid line shows that when the leading edge component is relatively
  prominent, the absolute flux density of the profile peak is higher.}
\label{0738_bright}
\end{center}
\end{figure*}

PSR~J0738$-$4042 is a bright pulsar with rotational properties typical
of the main population of middle-aged, isolated radio
pulsars. Despite this, it is known to show a dramatic change in both
pulse profile and $\dot{\nu}$, beginning in 2005 and resulting in a
new profile component
\citep{2011MNRAS.415..251K,2014ApJ...780L..31B}. Regular Parkes
observations of the pulsar began in March 2008. Figure~\ref{0738_map}
shows that a prominent change in profile shape occurred in November
2010 ($\sim$ MJD 55525), when the relative size of the transient
component (around pulse phase 0.08) increased significantly, and has
shown a trend of gradual recession ever since. 
\\
Figure~\ref{0738_map} shows a pattern of systematic changes that last
hundreds of days. In addition, different regions of the profile appear
to vary in a correlated or anti-correlated manner. Comparing Panels A
and B, shows this to be a good example of a pulsar that exhibits a
combination of profile shape changes (Panel B) and total flux density
changes (Panel A). Panel C reveals high S/N variability in the
estimated $\dot{\nu}$ with what appears to be the periodic signature
of a residual error in the position of the pulsar. The value of
$\dot{\nu}$ does not display any unusual behaviour around MJD 55525
when the primary change in pulse profile shape occurs.
\\
Panel A of Figure~\ref{0738_profile} highlights the regions of the
profile that are most variable. Examples of two significantly
different profiles are shown in red and blue in Panel B. Panel C shows
a complicated relationship between $\dot{\nu}$ and the profile
changes, with highest correlation seen at a pulse phase of $\sim$
0.075, and a lag of around zero. The previously noted correlation and
anti-correlation between regions of the profile can also be seen in
Panel C, in the alternating vertical patches.
\\
To illustrate the connection between different regions of the profile,
in Figure~\ref{0738_bright} we show the average profile of all the
epochs where the transient component is bright and compare it to the
average of all remaining epochs. We see that the main component is
also brighter when the transient component is bright, and note a small
suppression at the leading edge.
\subsubsection{PSR~J0742$-$2822 (B0740$-$28)}
\begin{figure*}
\begin{center}
\includegraphics[width=150mm]{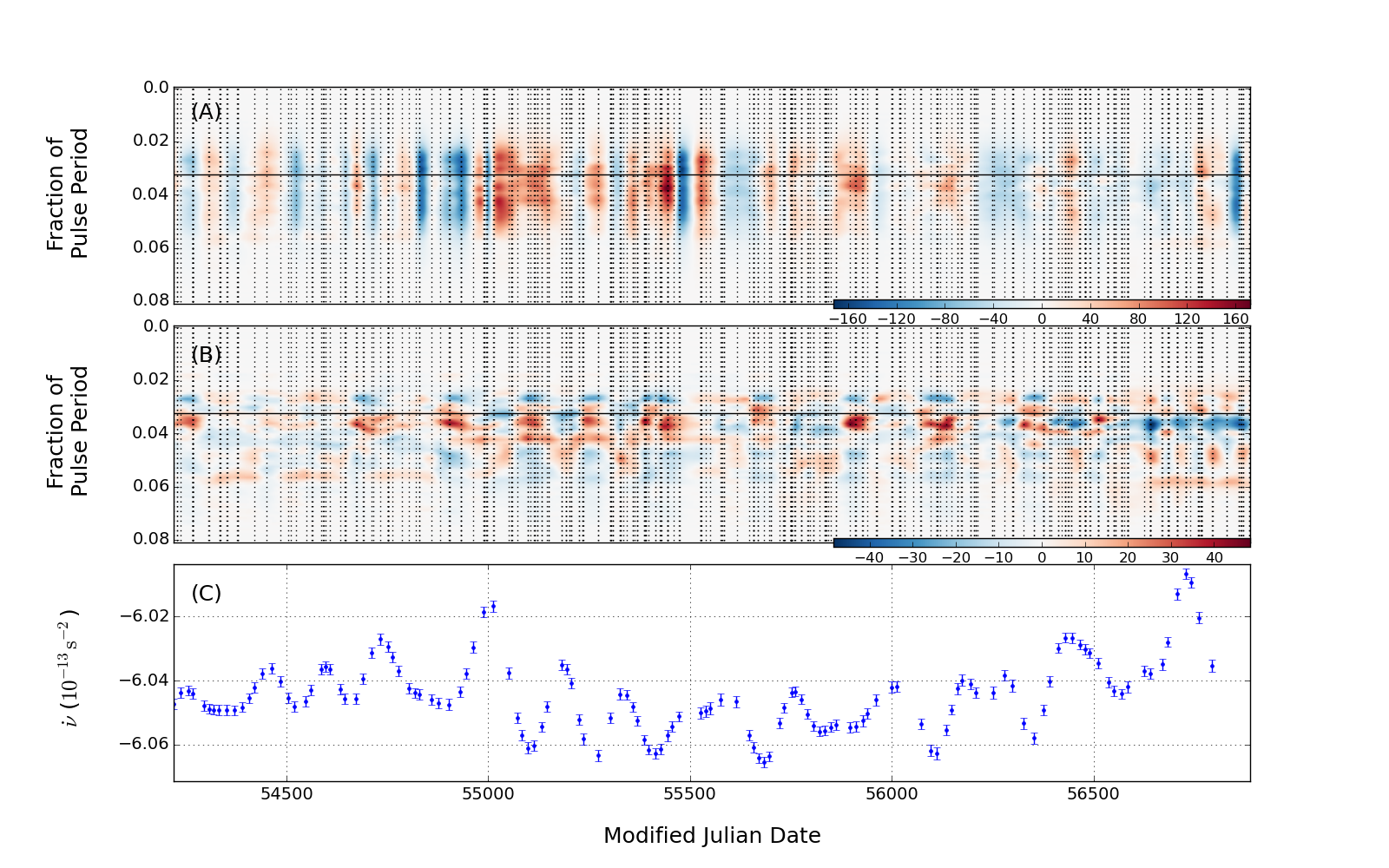}
\caption{Pulse profile and spindown variability for PSR~J0742$-$2822. As
  Figure~\ref{0738_map} otherwise.}
\label{0742_map}
\end{center}
\end{figure*}

\begin{figure*}
\begin{center}
\includegraphics[width=150mm]{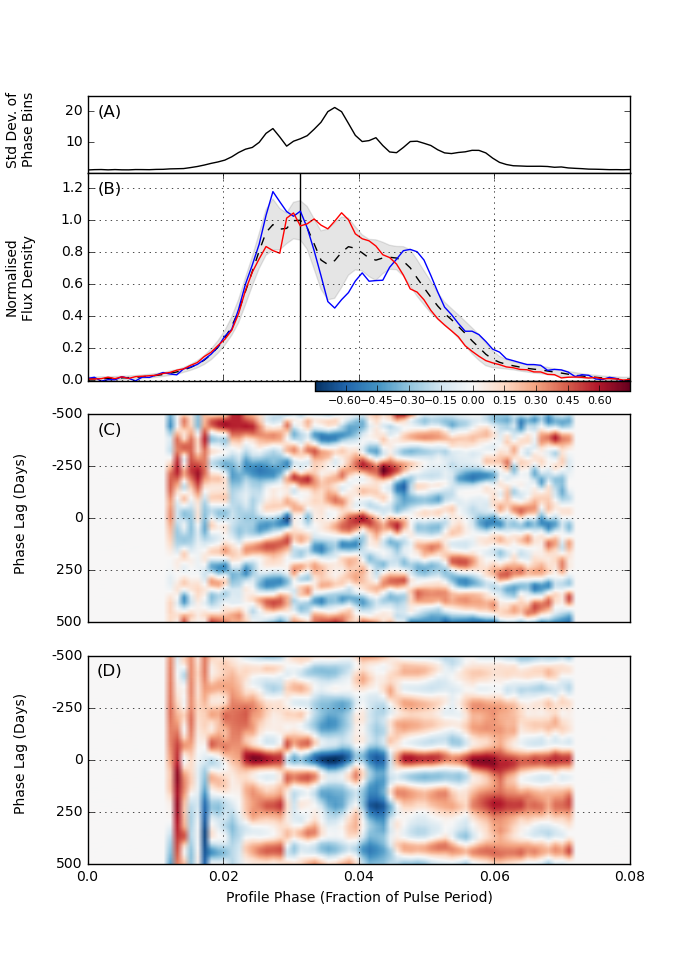}
\caption{Pulse profile variability and correlation map for
  PSR~J0742$-$2822. In Panel B, the blue profile was observed
on MJD 56642, red on MJD 55445. Panel C shows a correlation map composed of
data preceding a glitch on MJD 55022. Panel D is composed
only of data after the glitch. Otherwise as Figure~\ref{0738_profile}.}
\label{0742_profile_join}
\end{center}
\end{figure*}

PSR~J0742$-$2822 is known to display profile changes that correlate with
$\dot{\nu}$ variations \citep{2013MNRAS.432.3080K} and exhibits the
most rapid changes of the six state-switching pulsars analysed in
\citet{2010Sci...329..408L}.
\\
Panel A of Figure~\ref{0742_map} shows the flux density across the
whole pulse profile to vary. The individual observations show that the
flux density varies by $\sim$ 50\% of its median value. The rapid
changes in profile shape can be seen in Panel B of
Figure~\ref{0742_map}. This is most pronounced between MJD 55000 and
MJD 55500, where the changes in pulse profile can be seen, by eye, to
correlate with the $\dot{\nu}$ changes in Panel C of
Figure~\ref{0742_map}. The $\dot{\nu}$ timeseries is very similar to
the timeseries published by \citet{2013MNRAS.432.3080K}, using largely
the same dataset. As described previously, it is calculated without
enforcing a particular window and the error bars are computed using
all data that contribute to each point.
\\
Keith et al. summarise the profile changes by introducing a shape
parameter. They show that the shape parameter correlates particularly
well with $\dot{\nu}$ after a glitch (included in the timing model)
occurs at MJD 55022. Because of this, the disorganized pre-glitch
correlation map becomes organized post-glitch, where high correlation
is observed at zero lag. On closer inspection of Panel C of Figure~\ref{0742_profile_join}, $\dot{\nu}$
appears to be anti-correlated with the two profile peaks, and
correlated with the central trough. In contrast, in Panel D,
$\dot{\nu}$ appears to be correlated with the peaks, and
anti-correlated with the trough. It follows, and can be seen, that
certain phases of the pulse profile are correlated and anti-correlated
with others. This phenomenon can also be seen clearly in
PSR~J0738$-$4042, PSR~J1830$-$1059 and PSR~J1602$-$5100.
\\
Panel A of Figure~\ref{0742_profile_join} shows that the variability
across the profile does not follow the profile shape; the maximum
variability is near the centre of the profile as show in the examples
of Panel B.
\subsubsection{PSR~J1830$-$1059 (B1828$-$11)}
\begin{figure*}
\begin{center}
\includegraphics[width=150mm]{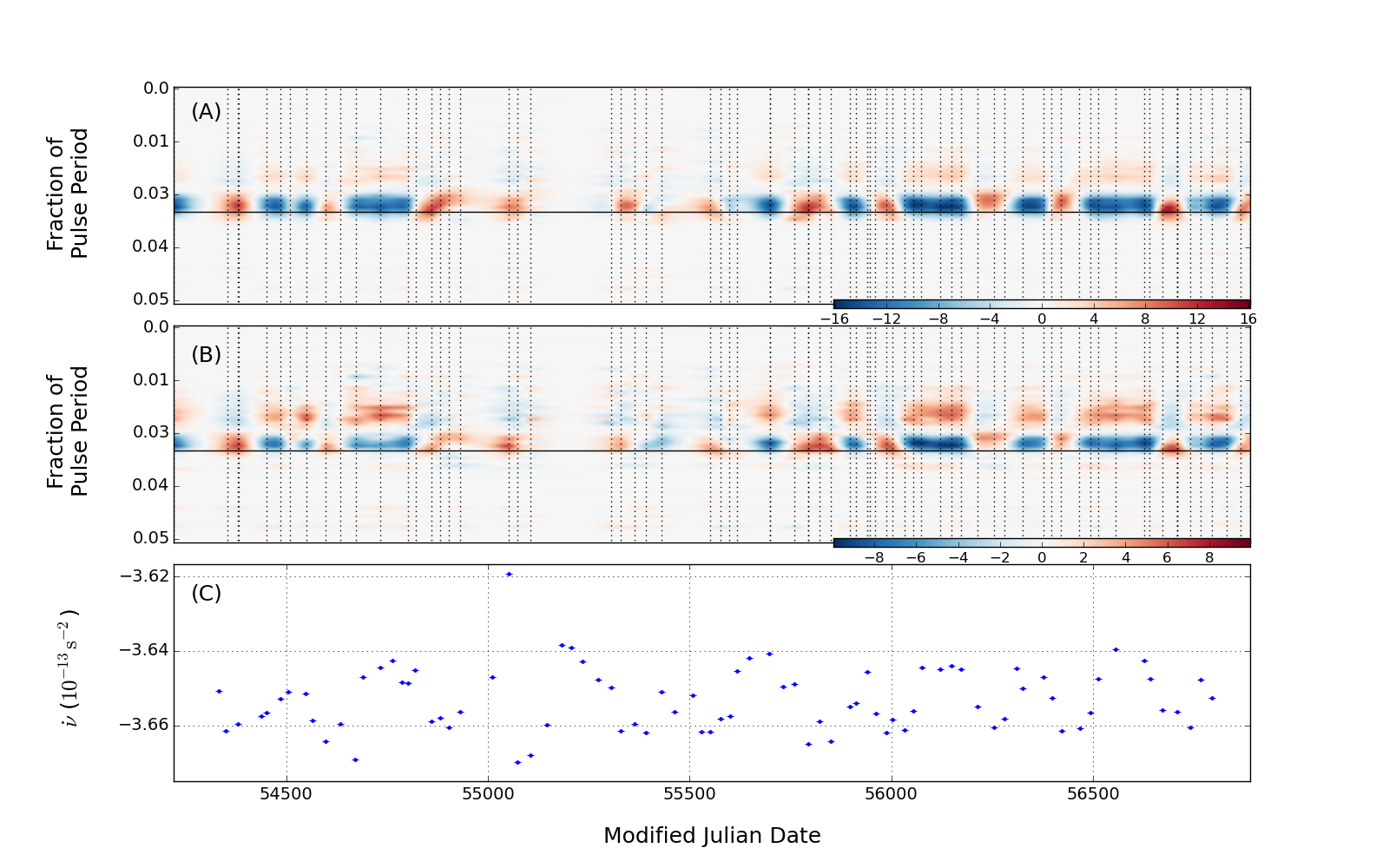}
\caption{Pulse profile and spindown variability for PSR~J1830$-$1059. As
  Figure~\ref{0738_map} otherwise.}
\label{1830_map}
\end{center}
\end{figure*}

\begin{figure*}
\begin{center}
\includegraphics[width=150mm]{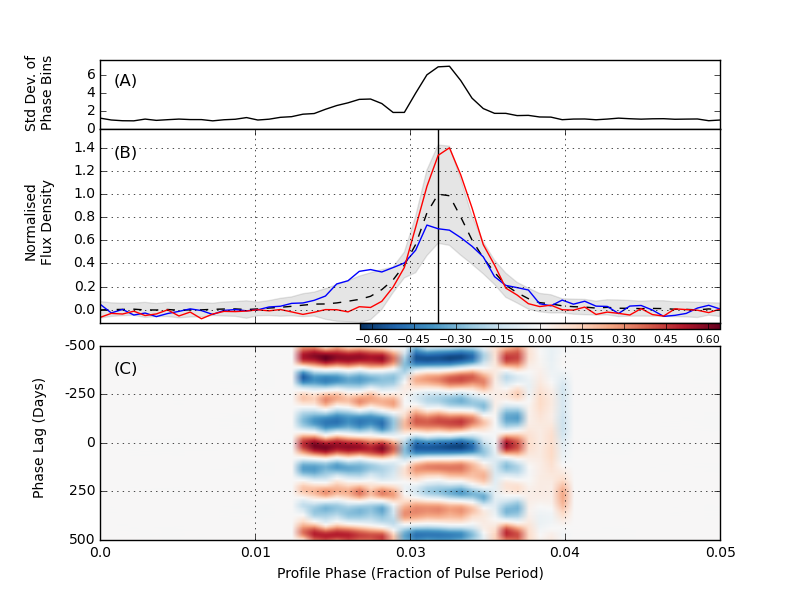}
\caption{Pulse profile variability and correlation map for
PSR~J1830$-$1059. In Panel B, the blue profile was observed
on MJD 54548, red on MJD 54381. Otherwise as
Figure~\ref{0738_profile}.}
\label{1830_profile}
\end{center}
\end{figure*}

\begin{figure*}
\begin{center}
\includegraphics[width=140mm]{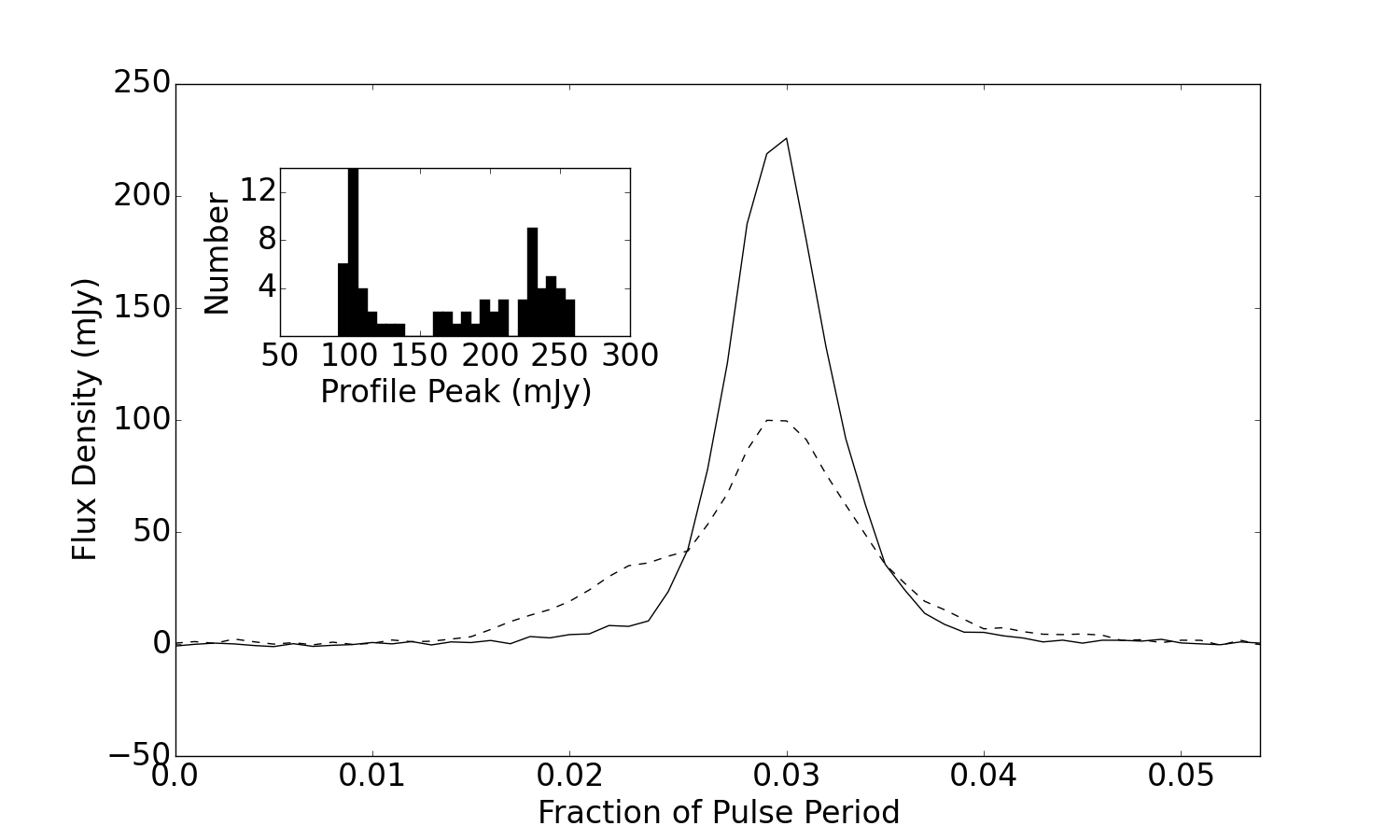}
\caption{Links between the profile shape and flux density of PSR~J1830$-$1059. The
  solid line traces the median of pulse profile that have a peak flux
  density greater than 150 mJy. The
  dashed line traces the median of pulse profiles that have a peak flux
  density lower than 150 mJy. The profiles with the highest flux density are also seen to be the most
narrow. The inset shows the flux density level of the profile peaks to be
bimodal, reflecting the two states.}
\label{1830_bright}
\end{center}
\end{figure*}

Long-term variability in both the pulse profile shape and spindown
rate are well established in PSR~J1830$-$1059, along with correlation
between the two \citep{2010Sci...329..408L}. The variability has been
attributed to free precession \citep{2000Natur.406..484S,2012MNRAS.420.2325J} or the
effects of an orbiting quark planet
\citep{2007MNRAS.381L...1L}. Quasi-periodic profile changes can be
seen clearly in Panels A and B of Figure~\ref{1830_map}. By eye, these
both seem to correlate with $\dot{\nu}$ in Panel C.
\\
The two example profiles in Panel B of Figure~\ref{1830_profile}
illustrate the differences in profile shape, where power from the
leading edge migrates to the pulse peak, and vice versa. Panel A shows
that the region just before pulse phase 0.03 remains relatively
constant.
\\
Figure~\ref{1830_bright} demonstrates that the two profile states
have different flux density levels; one has a peak flux density that
is more than a factor of two greater than that of the other. The more
intense state appears to have a simple profile composed of only one
component, whereas the dimmer state has a leading edge component. The
mean flux density of the on-pulse region of the two states is not
equal; the narrow state has a mean flux density around 1.4 times
larger than the wide state.
\\
As a result of the relationship between flux density and profile
shape, we see in Figure~\ref{1830_map} that the variations in the
Panel A also seem to be correlated with those in Panels B and C. This
suggests that the observed changes in flux density for PSR~J1830$-$1059
are the result of intrinsic processes, rather than propagation
effects.
\\
Panel C of Figure~\ref{1830_profile} shows the correlation between
$\dot{\nu}$ and profile shape at each pulse phase, as a function of
the lag between the two timeseries. As the pulse profile shape is
known to vary in synchronisation with $\dot{\nu}$, we see the
strongest correlation occurring around zero lag. We also see that the
timseries are correlated when $\dot{\nu}$ has a $\pm$ 500 day
lag. This is due to the fact that $\dot{\nu}$ has a cycle of around
500 days, as seen in Panel C of Figure~\ref{1830_map}.
\\
It is worth noting that the $\dot{\nu}$ variability shows the same
pattern of major and minor peaks, as published in
\citet{2010Sci...329..408L}.
\subsection{New Variable Pulsars}
\subsubsection{PSR~J0908$-$4913 (B0906$-$49)}
\begin{figure*}
\centering
  \begin{tabular}{@{}c@{}}
    \includegraphics[width=150mm]{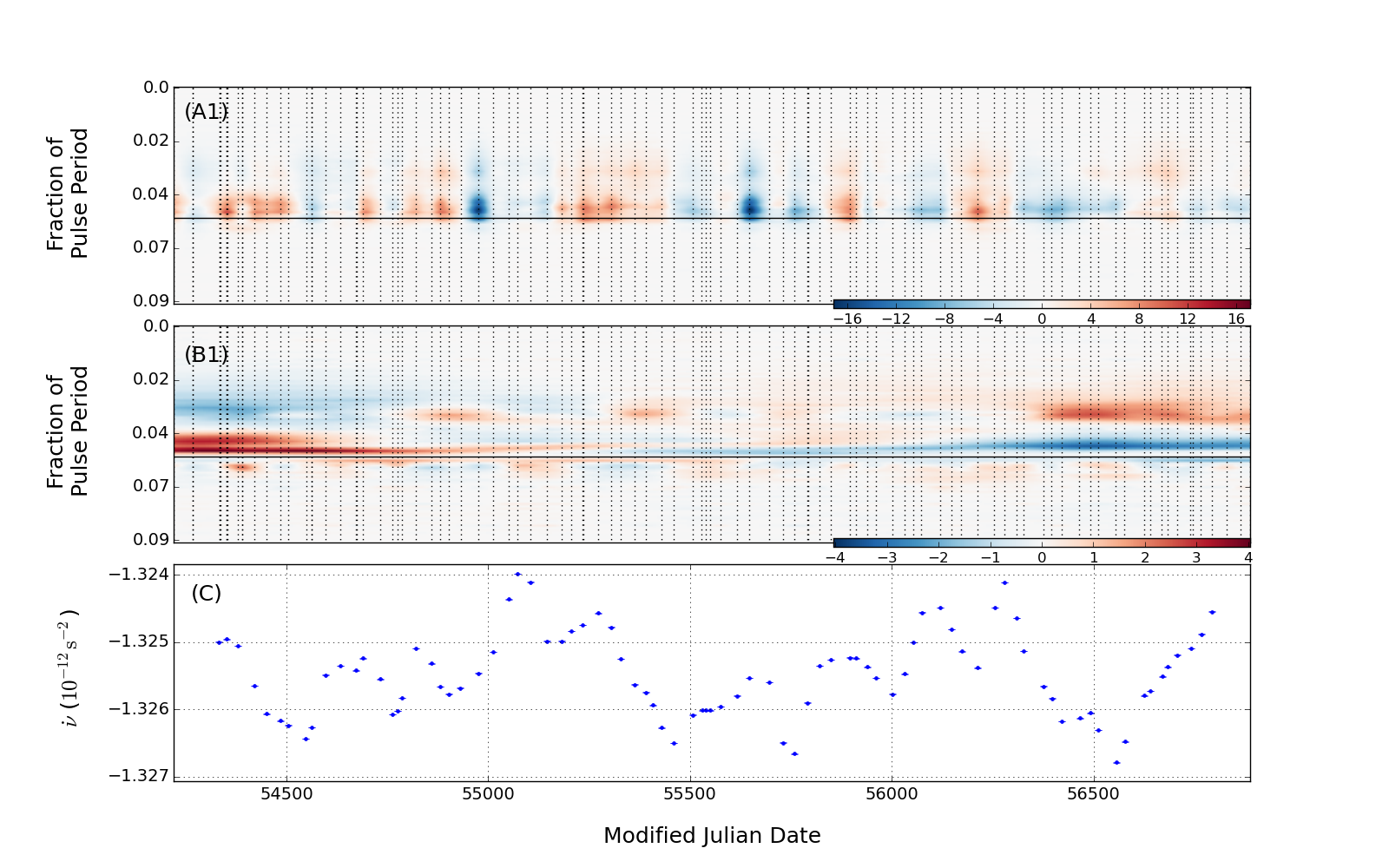}   \\
    \includegraphics[width=150mm]{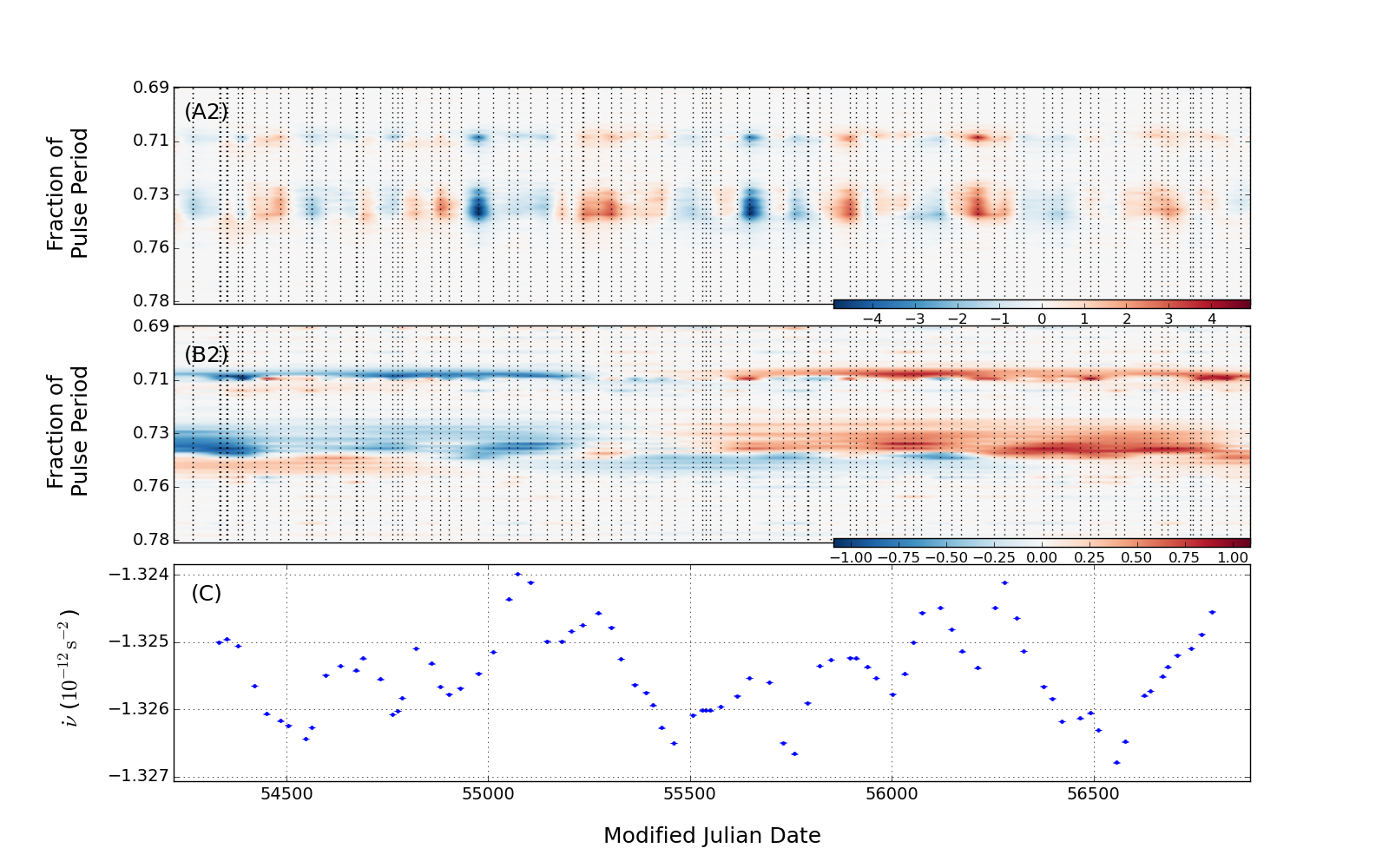}
  \end{tabular}
  \caption{Pulse profile and spindown variability for the MP and IP of
  PSR~J0908$-$4913. Panels A1 and B1 relate to the MP, while Panels A2
  and B2 relate to the and IP. As Figure~\ref{0738_map} otherwise.}
\label{0908_map}
\end{figure*}

\begin{figure*}
\centering
  \begin{tabular}{@{}c@{}}
    \includegraphics[width=130mm]{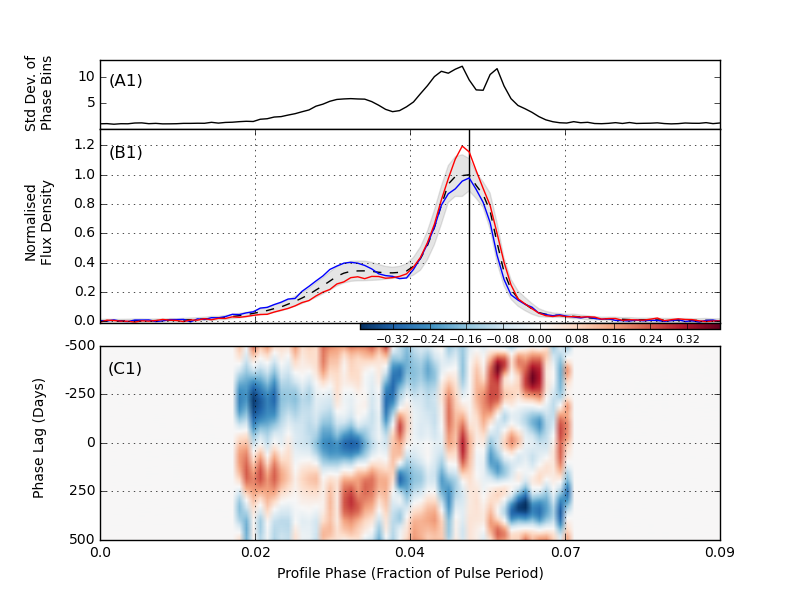}   \\
    \includegraphics[width=130mm]{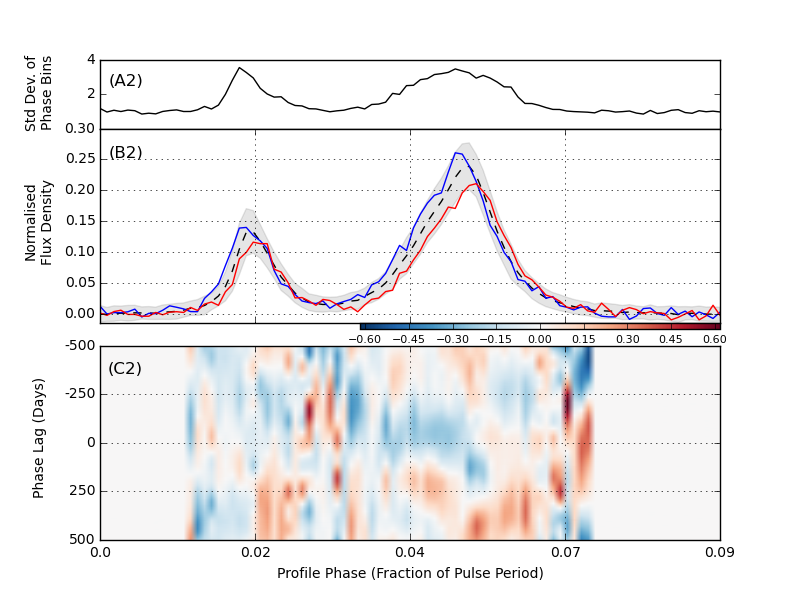}
  \end{tabular}
  \caption{Pulse profile variability and correlation map for the MP
    and IP of PSR~J0908$-$4913. Panels A1, B1 and C1 relate to the MP,
    while Panels A2, B2 and C2 relate to the IP. In Panel B1 and B2, the blue profile was observed
on MJD 56670, red on MJD 54350. Otherwise as Figure~\ref{0738_profile}.}
\label{0908_profile}
\end{figure*}

The profile of PSR~J0908$-$4913 consists of a main pulse (MP) and an
interpulse (IP), and has been show by
\cite{2008MNRAS.390...87K} to be an orthogonal rotator. Profile-wide
flux density variations can be seen in both the MP and IP (Panels A1
and A2 of Figure~\ref{0908_map} respectively); the emission received
from PSR~J0908$-$4913 varies quasi-periodically across the dataset. The
flux calibrated profiles vary by up to $\sim$ 50\% from the median,
which has a peak of around 1000 mJy. The changes in the shape of the
pulse profile (Panels B1 and B2 of Figure~\ref{0908_map}), however,
are slight and gradual across the dataset; both IP components and a
precursor to the MP steadily grow with respect to the MP.
In Panel C, $\dot{\nu}$ shows quasi-periodicity. We see slow
variability with three local minima around 54500, 55500 and 56500,
between which, we see around four shorter-term cycles. There is no
obvious correlation with the shape changes, as is confirmed by the
correlation maps in Figure~\ref{0908_profile}. Given the alignment
techniques used, the two example profiles in Panels B1 and B2 of
Figure~\ref{0908_profile} show what appears to be a shift in phase in
the IP, which is not seen in the MP.
\subsubsection{PSR~J0940$-$5428}
\begin{figure*}
\begin{center}
\includegraphics[width=150mm]{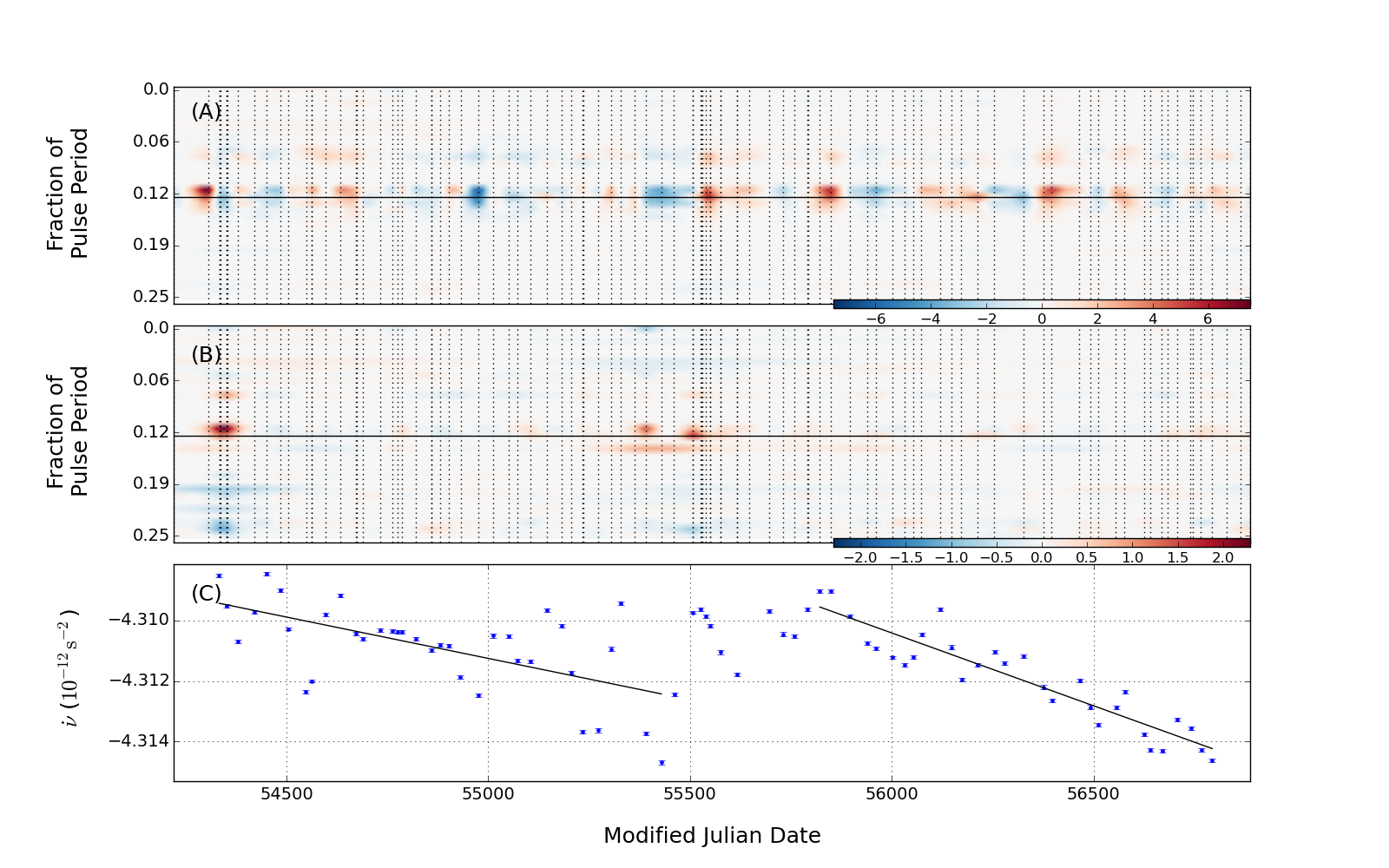}
\caption{Pulse profile and spindown variability for
  PSR~J0940$-$5428. The black sold lines in Panel C are best fit to the data points that they span. The left gradient is $-2.73\times 10^{-18}$ s$^{-2}$day$^{-1}$
and the right gradient is $-4.80\times 10^{-18}$ s$^{-2}$day$^{-1}$. As Figure~\ref{0738_map} otherwise.}
\label{0940_map}
\end{center}
\end{figure*}

\begin{figure*}
\begin{center}
\includegraphics[width=150mm]{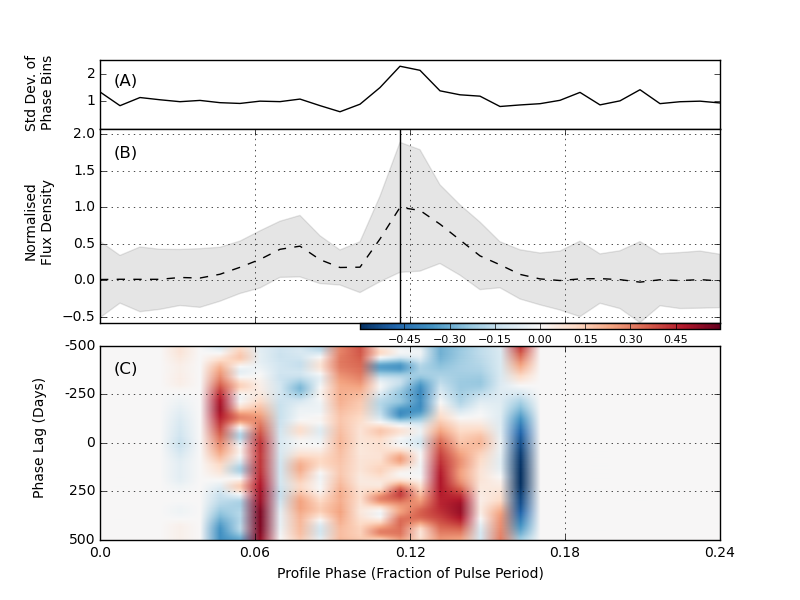}
\caption{Pulse profile variability and correlation map for
  PSR~J0940$-$5428. As Figure~\ref{0738_profile} except there are no red
  and blue traces representing pulse profile changes, as they are
  minimal. This number of phase bins for this pulsar has been reduced
  to 128.}
\label{0940_profile}
\end{center}
\end{figure*}

PSR~J0940$-$5428 is a faint pulsar; the peak of the median profile is
$\sim$ 12 mJy. The pulsar shows only low significance
variations in flux density and profile shape (Panels A and B of
Figure~\ref{0940_map} respectively). We feature PSR~J0940$-$5428, however,
as an example of $\dot{\nu}$ changes in the absence of significant
pulse profile variability.
Panel C of Figure~\ref{0940_map}, shows systematic changes of
$\dot{\nu}$ on a time-scale of $\sim$ 200 days, and also two
longer-term trends, the gradients of which are marked by solid
lines. A steady $\dot{\nu}$ gradient is indicative of a constant
$\ddot{\nu}$ term that is absent in the timing model. From left to
right, the two $\ddot{\nu}$ terms are $-2.73\times 10^{-18}$ s$^{-2}$day$^{-1}$
and $-4.80\times 10^{-18}$ s$^{-2}$day$^{-1}$.
\\
The lack of pulse profile variability is reflected by Panel A and
Panel B of Figure~\ref{0940_profile}, in which the deviations of the
on-pulse phase bins are comparable to the off-pulse deviations across
the profile. As expected, the correlation plot in Panel C of
Figure~\ref{0940_profile} does not show any significant structure.
\subsubsection{PSR~J1105$-$6107}
\begin{figure*}
\begin{center}
\includegraphics[width=150mm]{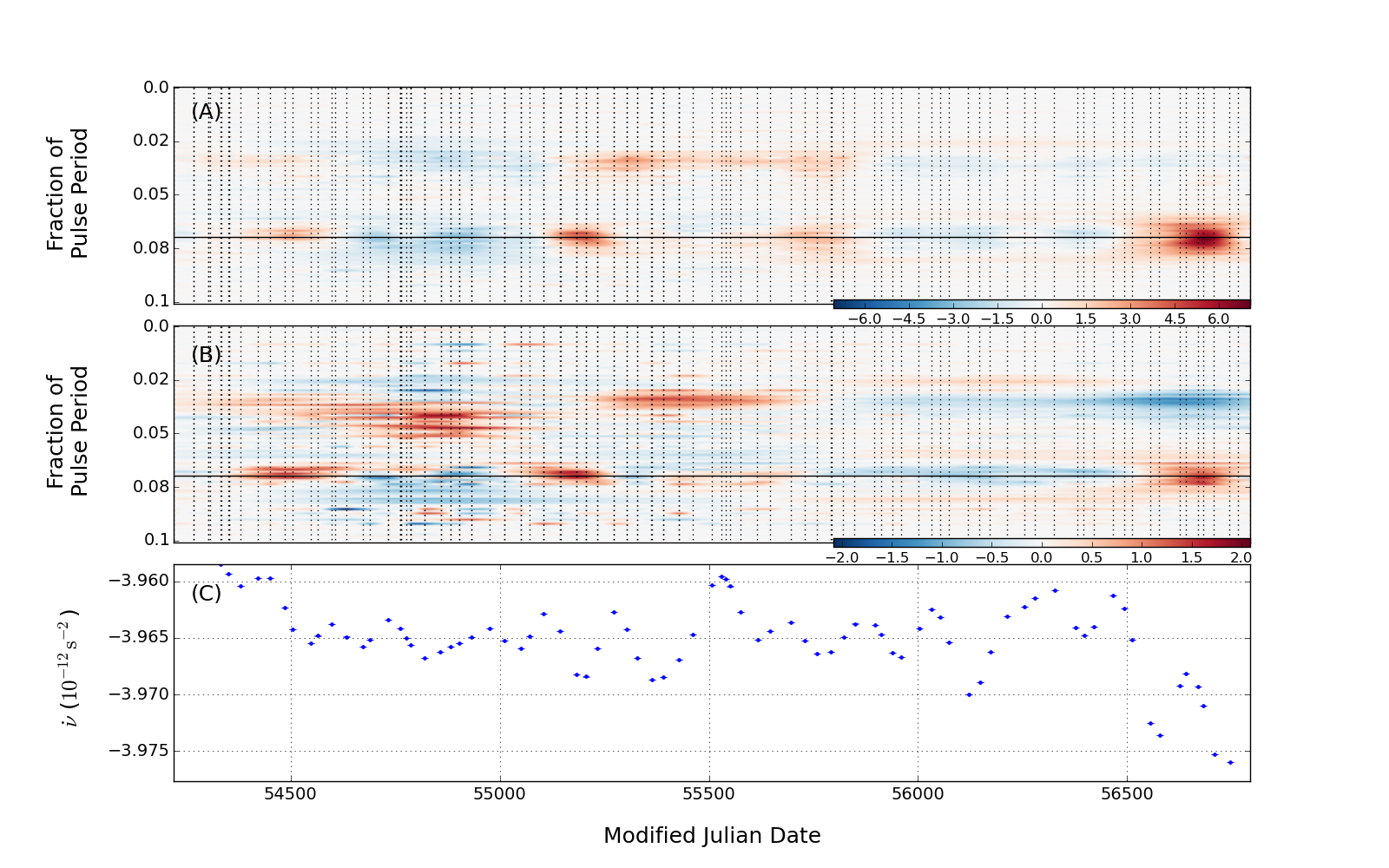}
\caption{Pulse profile and spindown variability for PSR~J1105$-$6107. As
  Figure~\ref{0738_map} otherwise.}
\label{1105_map}
\end{center}
\end{figure*}

\begin{figure*}
\begin{center}
\includegraphics[width=150mm]{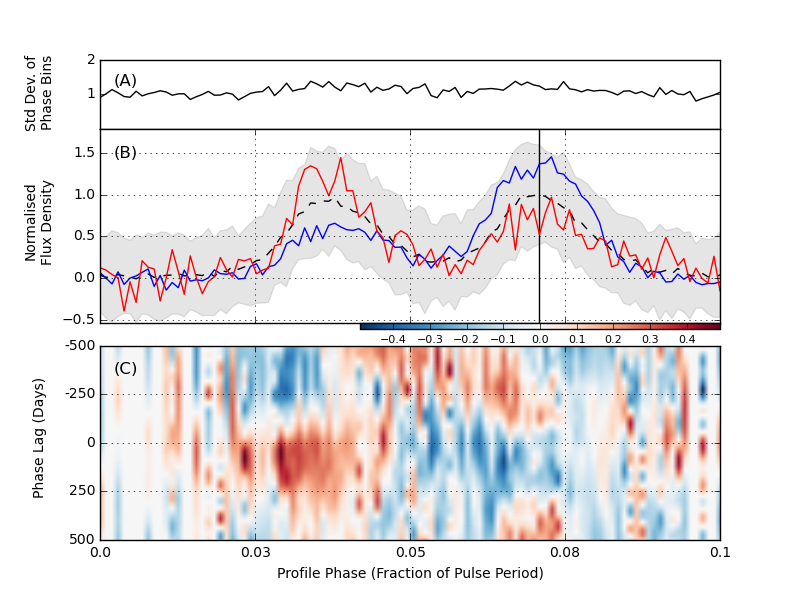}
\caption{Pulse profile variability and correlation map for
  PSR~J1105$-$6107. In Panel B, the blue profile was observed
on MJD 56640, red on MJD 55304. Otherwise as Figure~\ref{0738_profile}.}
\label{1105_profile}
\end{center}
\end{figure*}

PSR~J1105$-$6107 has a spin period of 63 ms and characteristic age of
only 63 kyr. The pulsar has a possible association with a nearby
supernova remnant \citep{1997ApJ...485..820K}. Both Panel A and Panel
B of Figure~\ref{1105_map} highlight systematic profile
variations. One significant shape change (in the relative size of the
two profile components) occurs over several observations, beginning
$\sim$ MJD 56500.  This change coincides with an increase in
$\dot{\nu}$, seen in Panel C. Aside from this, any shape changes are
subtle and the profile shape is largely stable across the
dataset. This is demonstrated in Panel A of Figure~\ref{1105_profile}; the level of on- and off-pulse profile deviation
is approximately the same.
\\
The nature and degree of the profile changes is shown in Panel B of
Figure~\ref{1105_profile}. Panel C shows that the highest SRCCs
between $\dot{\nu}$ and flux density variability occur around zero lag
in the leftmost of the two components.
\subsubsection{PSR~J1359$-$6038 (B1356$-$60)}
\begin{figure*}
\begin{center}
\includegraphics[width=150mm]{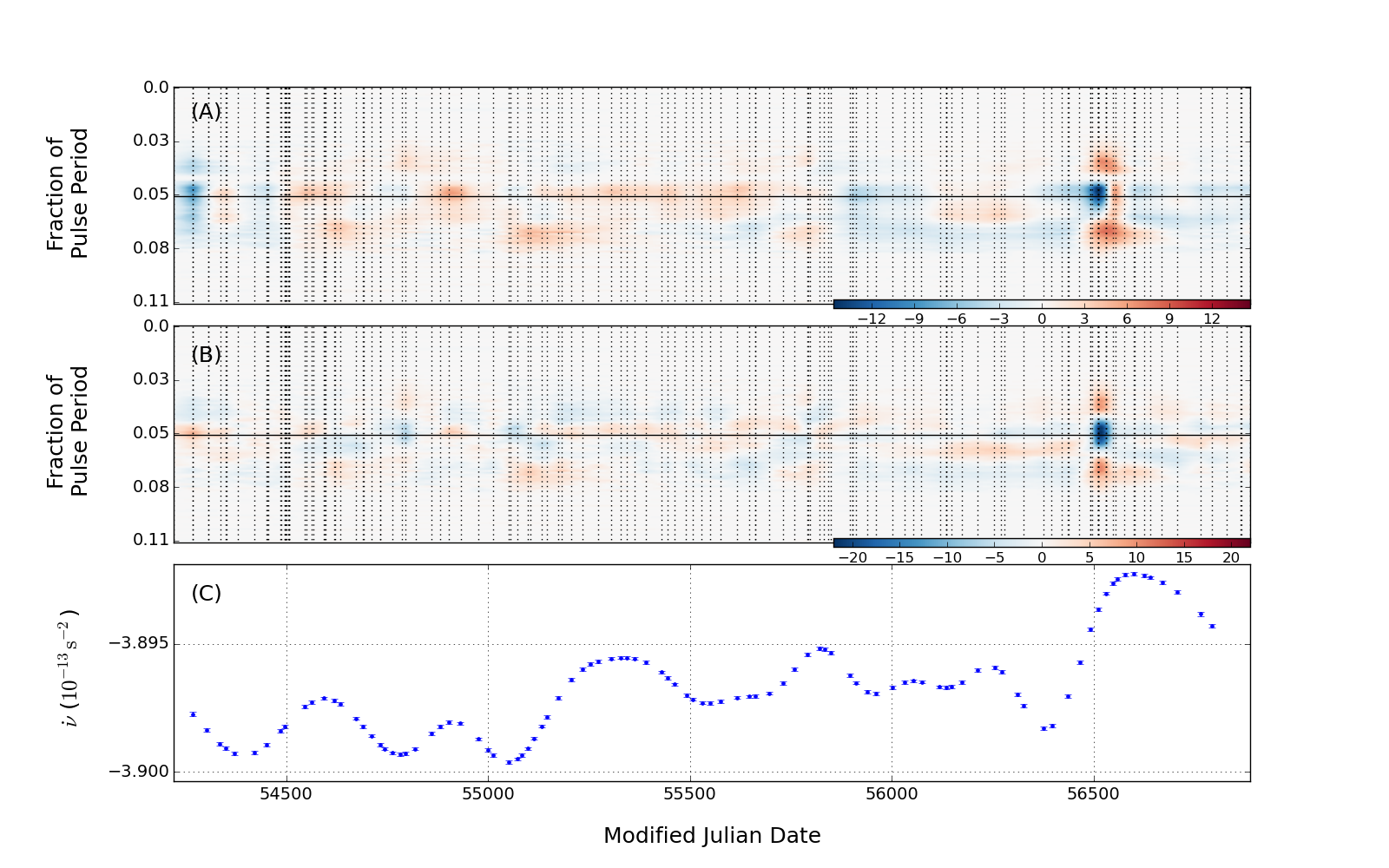}
\caption{Pulse profile and spindown variability for PSR~J1359$-$6038. As
  Figure~\ref{0738_map} otherwise.}
\label{1359_map}
\end{center}
\end{figure*}

\begin{figure*}
\begin{center}
\includegraphics[width=150mm]{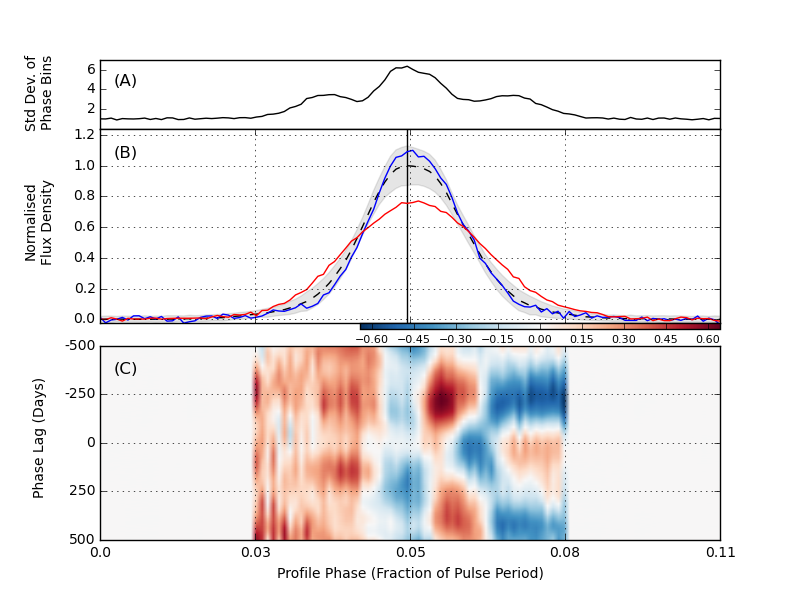}
\caption{Pulse profile variability and correlation map for
  PSR~J1359$-$6038. In Panel B, the blue profile was observed
on MJD 54268, red on MJD 56512. Otherwise as Figure~\ref{0738_profile}.}
\label{1359_profile}
\end{center}
\end{figure*}

The flux density of this pulsar systematically varies by $\sim$ 10\%
around the median. Panels A and B of Figure~\ref{1359_profile} indicate that there is some profile
variability throughout the dataset, but the most significant shape
changes by far occur on three observation days: MJD 56512, 56513 and 56531
(Panel B of Figure~\ref{1359_map}). The normalised pulse profiles of
these observations are substantially wider and shorter than the median
profile. Panel C shows that this change in pulse shape approximately coincides with a
drop in spindown rate.
\\
Panel A of Figure~\ref{1359_profile} demonstrates that most
variability is occurring at the peak and in the wings of the pulse
profile. Panel B shows examples of extreme profile shapes; the red
profile reflects the shape of the three anomalous observations
discussed above. This profile is mostly outside the grey 2$\sigma$
bands of the median profile, confirming that these short-term profile
shape changes exceed the typical level of profile variability across
the dataset. In Panel C of Figure~\ref{1359_profile}, the highest
level of correlation is seen when the $\dot{\nu}$ variability lags the
flux density variability by 250 days (a phase lag of -250 days).
\subsubsection{PSR~J1600$-$5044 (B1557$-$50)}
\begin{figure*}
\begin{center}
\includegraphics[width=150mm]{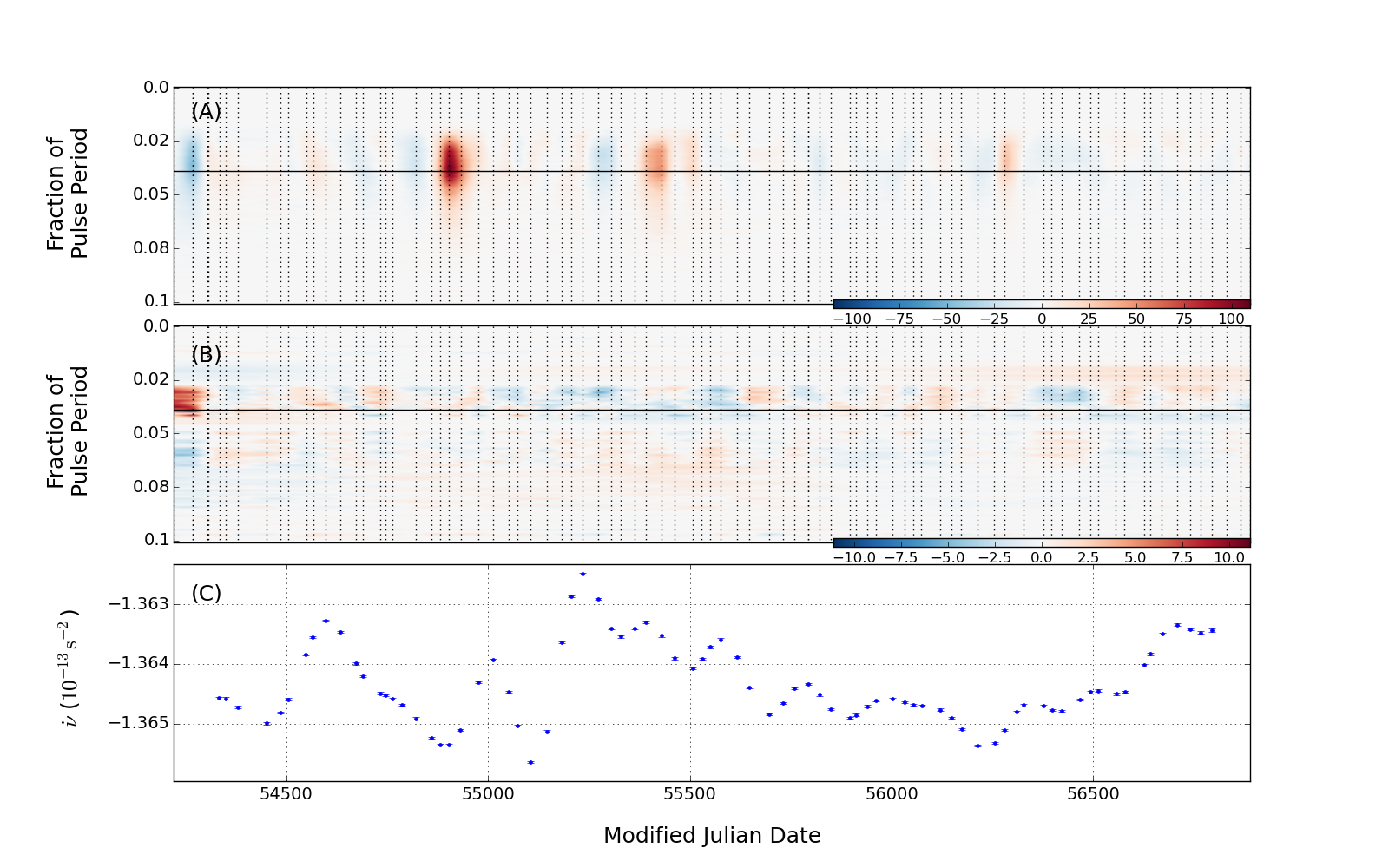}
\caption{Pulse profile and spindown variability for PSR~J1600$-$5044. As
  Figure~\ref{0738_map} otherwise.}
\label{1600_map}
\end{center}
\end{figure*}

\begin{figure*}
\begin{center}
\includegraphics[width=150mm]{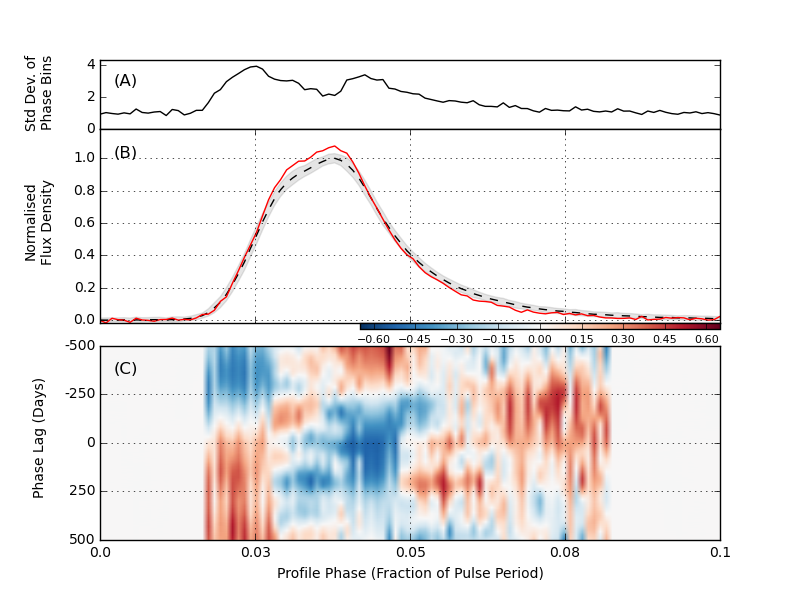}
\caption{Pulse profile variability and correlation map for
  PSR~J1600$-$5044. The red profile was observed on MJD. This profile
  shape is seen in the only two observations with noticeable deviations
  from the median. As Figure~\ref{0738_profile} otherwise}
\label{1600_profile}
\end{center}
\end{figure*}

Observations of this pulsars over a 13 year dataset from Hartebeesthoek Radio Astronomy
Observatory showed evidence of cyclic variations in both the
dispersion measure and $\dot{\nu}$ of PSR~J1600$-$5044
\citep{2002AfrSk...7...41C}. The two appear anti-correlated, and are
attributed to free precession of the pulsar.
\\
In the Parkes dataset, flux calibrated observations of PSR~J1600$-$5044
show significant variability; the peak of the median profile is $\sim$
800 mJy, which increases systematically to $\sim$ 1500 mJy (MJD 54902)
before dropping back to the level of the median value. The process
occurs over $\sim$ 120 days and can be seen in Panel A of
Figure~\ref{1600_map}). These flux density variations are expected to
be largely due to refractive scintillation, and are eliminated by the
normalisation process. Panel B shows that no significant variability
is seen in pulse profile shape, with the exception of the first two
observations of the dataset.  Slow $\dot{\nu}$ variability is seen in
Panel C, between which, shorter-term cycles are observed.
\\
In Figure~\ref{1600_profile}, Panel A shows that the most variability
occurs at the leading and trailing edges of the pulse profile; the
median value is shown in Panel B. The red trace in this panel
represents the profile shape of the first two observations in the
dataset. The narrow nature of the grey 2$\sigma$ bands demonstrates
the relative stability of the profile. The correlation plot in Panel C
of Figure~\ref{1600_profile} does not show significant structure. This
is expected, due to the lack of profile shape variation in the
dataset. 
\subsubsection{PSR~J1602$-$5100 (B1558$-$50)}
\begin{figure*}
\begin{center}
\includegraphics[width=150mm]{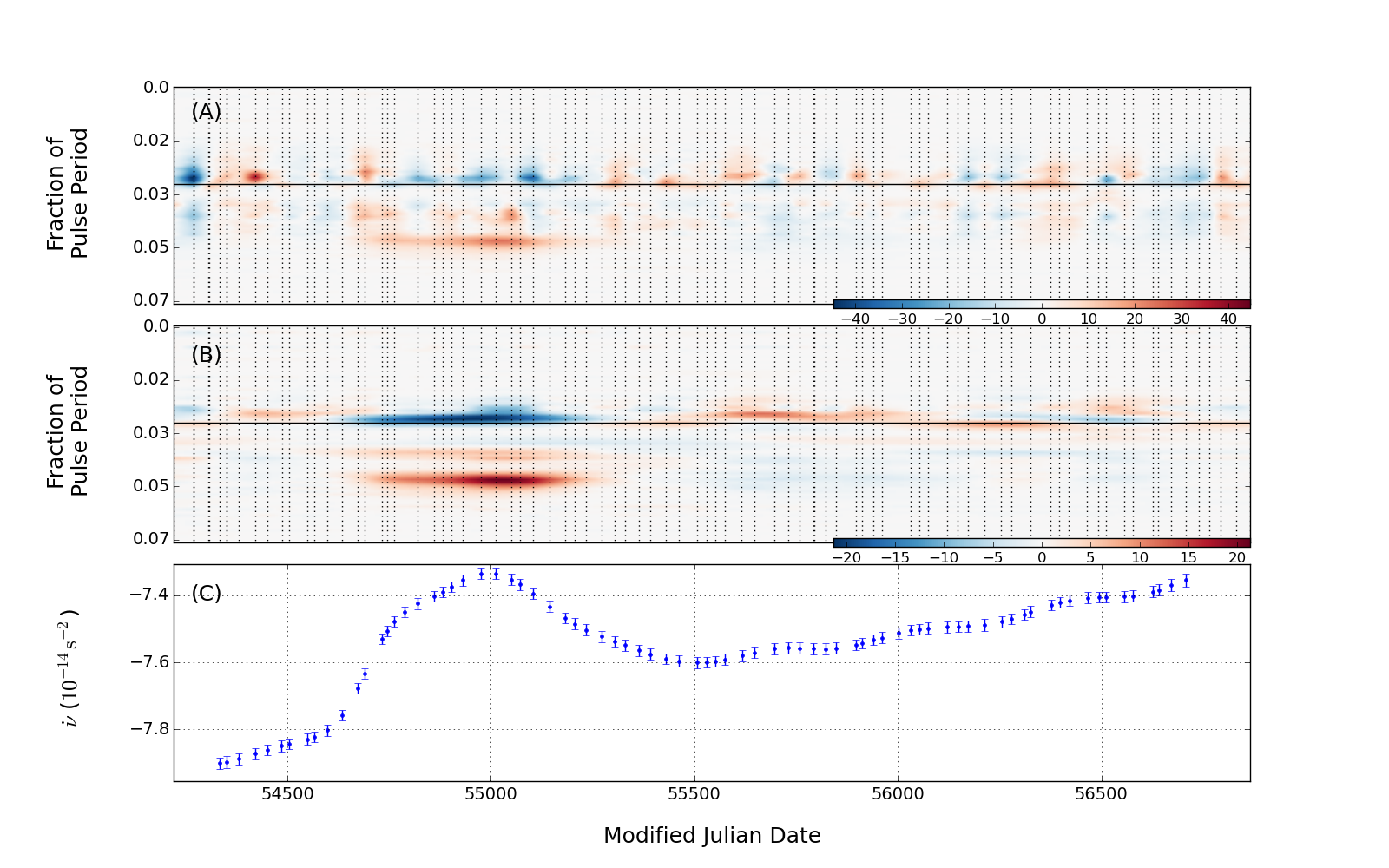}
\caption{Pulse profile and spindown variability for PSR~ J1602$-$5100. As Figure~\ref{0738_map} otherwise.}
\label{1602_map}
\end{center}
\end{figure*}

\begin{figure*}
\begin{center}
\includegraphics[width=150mm]{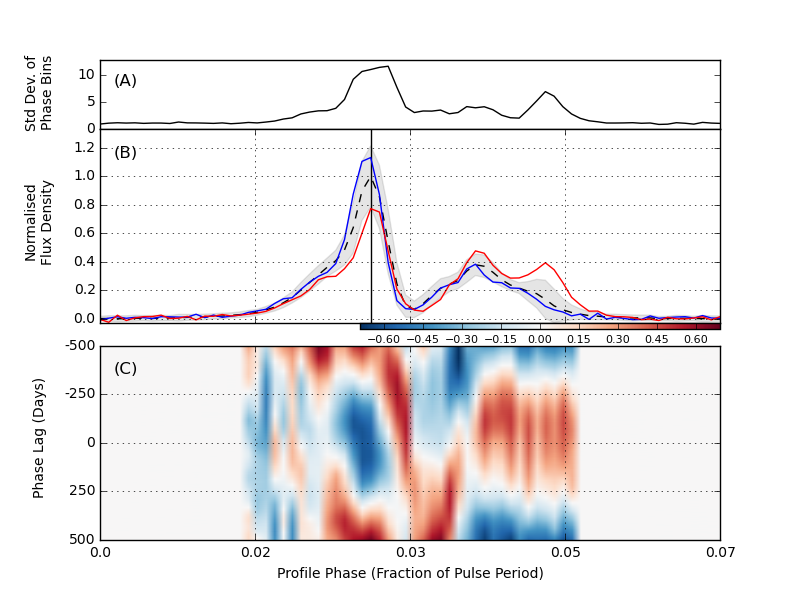}
\caption{Pulse profile variability and correlation map for
  PSR~J1602$-$5100. In the middle panel, the blue profile was observed
on MJD 54420, red on MJD 55072. Otherwise as Figure~\ref{1830_profile}.}
\label{1602_profile}
\end{center}
\end{figure*}

\begin{figure*}
\begin{center}
\includegraphics[width=140mm]{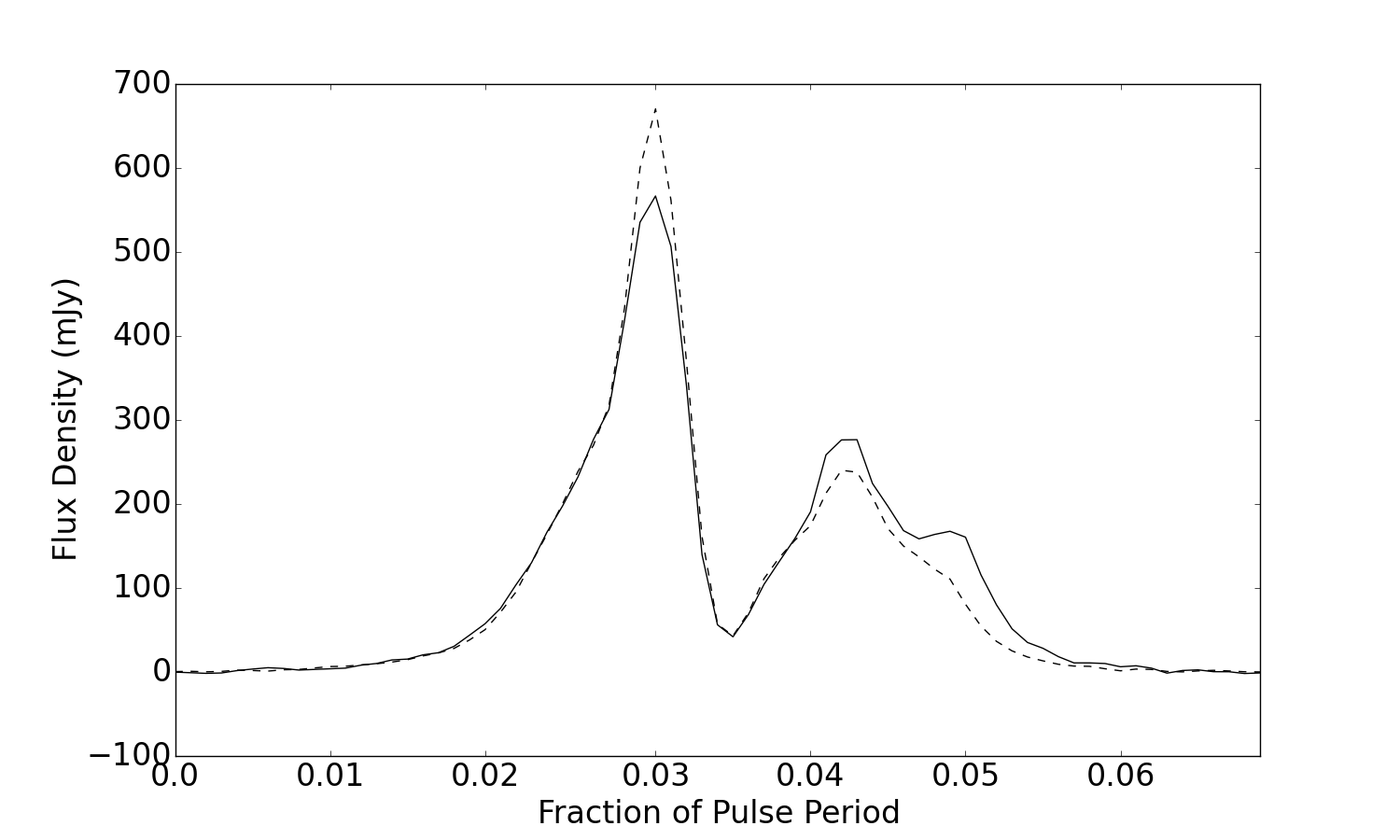}
\caption{Links between the profile shape and the flux density of PSR~J1602$-$5100. The
  solid line traces the median of pulse profiles that were observed
  between MJD 54672 and MJD 55304, i.e. the epoch over which the
  transient component appeared. The
  dashed line traces the median of the pulse profiles that fall
  outside this epoch. When the transient trailing edge component is present, the
peak flux is seen to fall.}
\label{1602_bright}
\end{center}
\end{figure*}

PSR~J1602$-$5100 has the smallest spin frequency of the nine pulsars
featured in this work, and consequently only PSR~J0738$-$4042 has a
lower spindown luminosity $\dot{E}$. A dramatic change in the pulse
profile can be seen in both Panel A and Panel B of
Figure~\ref{1602_map}, beginning at $\sim$ MJD 54700 and occurring
over $\sim$ 600 days. A drop in $\dot{\nu}$ of $\sim$ 5\%, which is
correlated with the shape change, can be seen in Panel C;  both sets
of variations appear to begin, peak and end at approximately the same
time.
\\
Panel A of Figure~\ref{1602_profile} shows that the most variability
occurs in the profile peak, and in a trailing edge component. The
extent of the pulse profile shape variations can be seen in Panel
B. In the correlation map (Panel C), SRCC between $\dot{\nu}$ and flux
density variability is seen to be strongest around the pulse profile
phase at which the transient component appears. The rising baseline of
$\dot{\nu}$ (indicative of an unmodelled $\ddot{\nu}$ parameter) was
flattened prior to the correlation calculations.
\\
The flux calibrated observations show that the appearance of the new
peak at the trailing edge of the smaller profile component coincides
with a drop in flux density at the primary profile component
(Figure~\ref{1602_bright}).
\subsection{Other Pulsars in the Dataset}
\label{others}
\begin{figure*}
\centering
  \begin{tabular}{@{}c@{}}
    \includegraphics[width=150mm]{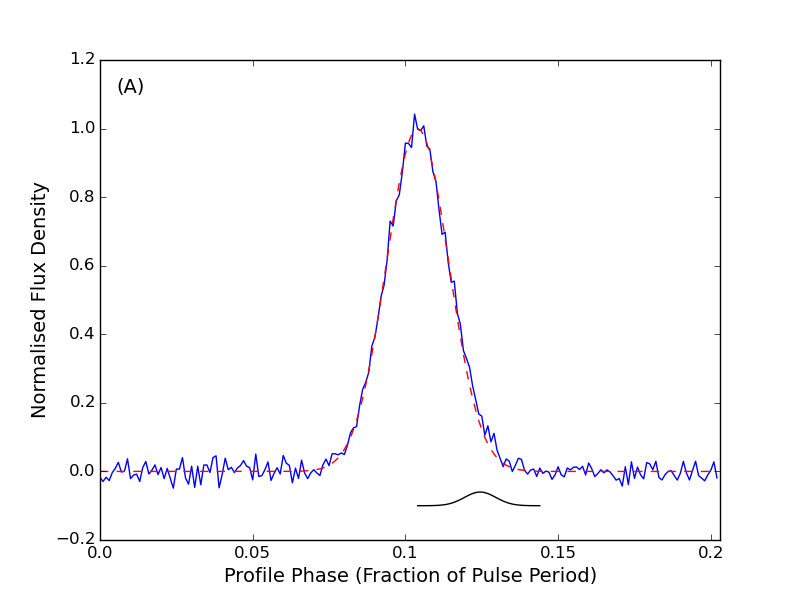}   \\
    \includegraphics[width=90mm]{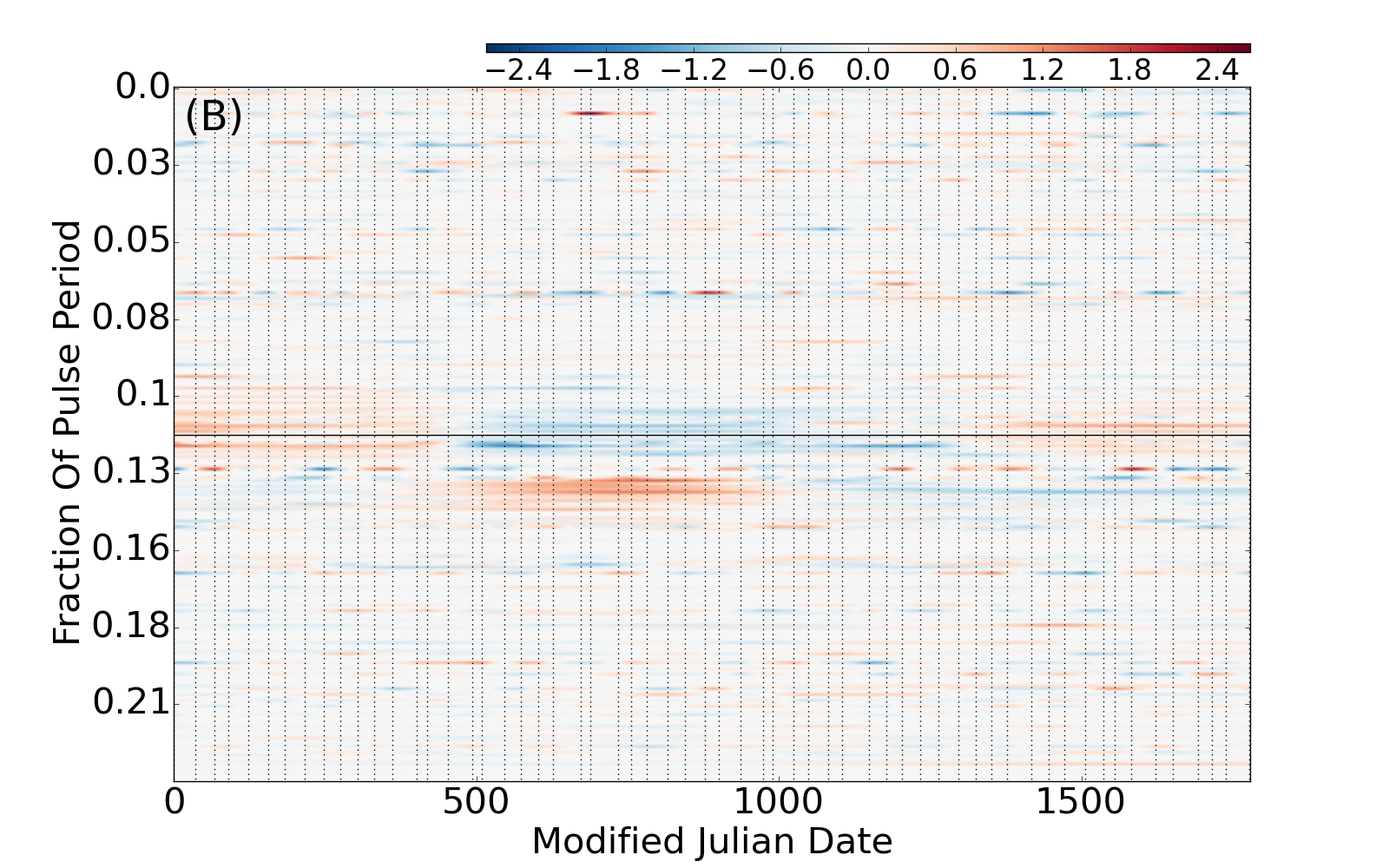}    
    \includegraphics[width=90mm]{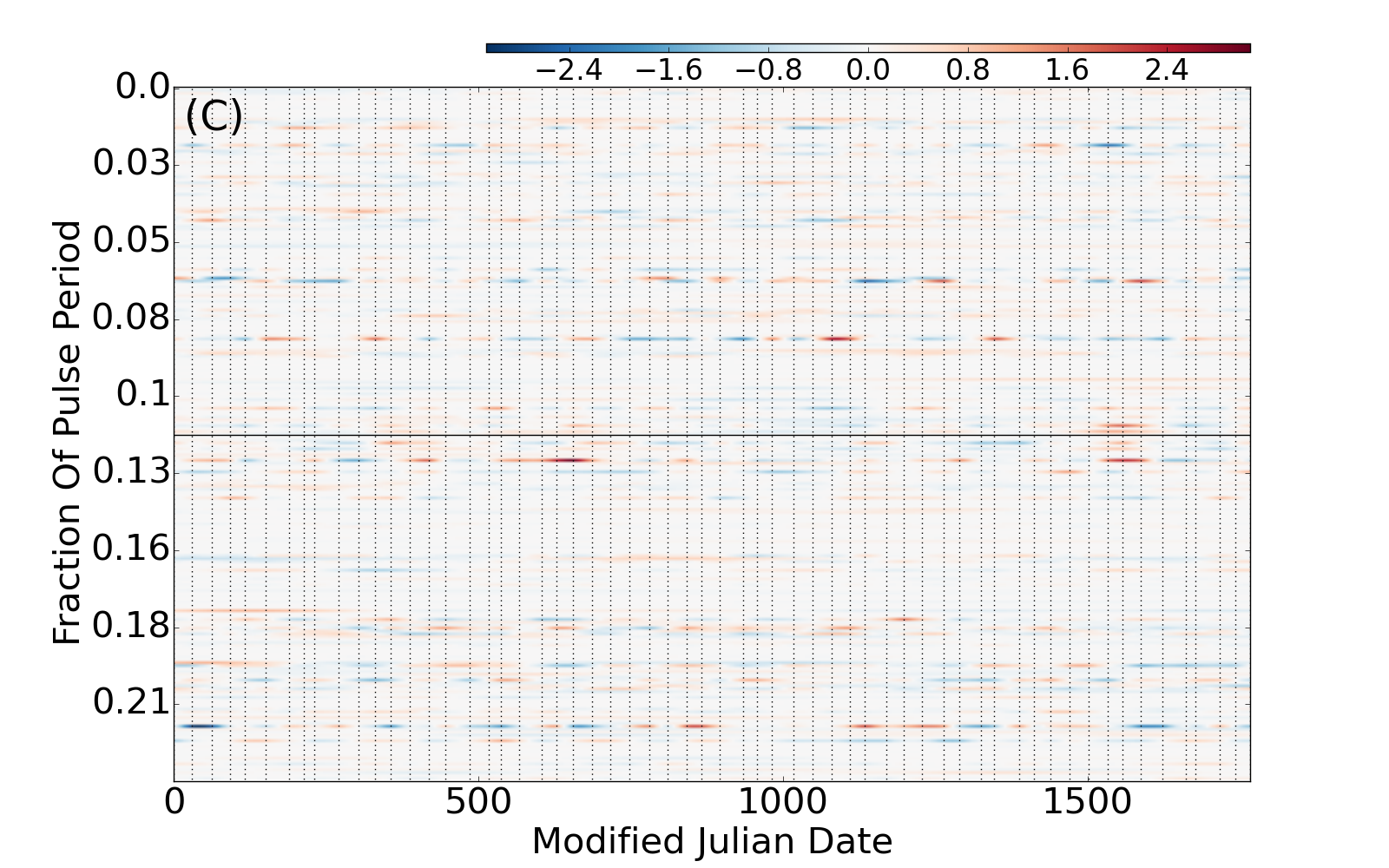}
  \end{tabular}
  \caption{The variability maps resulting from simulated pulse
    profiles. Over the $\sim$ 5 year simulation, a transient components grows
    and recedes over a $\sim$ 2.5 year span, on the trailing edge of the
    pulse profile. Panel A shows a comparison between the noiseless
    profile template (red dashed line) and the profile at the height
    of its deviation with noise included (blue solid line). The transient
    component is shown (at its most prominent) at the phase at which it appears, as a black
    solid line. Panels B and C show the difference between the
    simulated profiles and the profile template. The units are the
    median of the standard deviation of all off-pulse regions of the
    simulated profiles. The solid horizontal line highlights the
    profile peak, and the vertical dashed lines show the date of the
    simulated observations. Panel B is the variability map that results from this simulated
  dataset. Panel C shows the variability map that results
  after the simulated profiles are randomly shuffled in time.}
\label{sim}
\end{figure*}

The pulsars described above were chosen because they displayed some
type of noteworthy variability. The pulse profile variability maps of
all 168 pulsars in our dataset were assessed by eye. A non-detection
of variability results from a stable pulse profile, but also from
variations that are undetectable due to an insufficient S/N. The extent to
which variability can be detected in noisy profiles is discussed in
the next section.
\\
Performing the rotational variability analysis revealed that many of the
pulsars in our dataset displayed a yearly cycle in $\dot{\nu}$,
signifying a positional inaccuracy in the pulsar timing model. 
Rotational variability may, therefore, remain hidden in those pulsars.
We leave it for future work to investigate whether our modelling
technique can be used to better determine the position and proper
motion, as opposed to techniques involving the removal of timing noise
\citep{2011MNRAS.418..561C}.
\\
When analysing profile variability, only 9 out of the 168 pulsars
show significant profile shape changes. PSR~J1302$-$6350 and PSR J1825$-$0935
were not featured in this work but our analysis revealed their known
variability \citep{1994MNRAS.268..430J, 2010Sci...329..408L}. In order to
determine the type of profile variability that can pass undetected by
our analysis techniques, we produced a series of simulated pulsar
datasets. Artificial pulse profiles were created and spaced
approximately 30 days apart, spanning around 5 years in duration. The
pulse profile begins as a simple Gaussian function. After around a
year of the simulation, a small transient component grows and recedes
on the trailing edge of the profile over the course of 3 years of the
simulation. Noise with a standard deviation of 0.02 of the main pulse
peak was added to all profiles. The effects of refractive
scintillation were simulated by scaling each profile by a random
factor. The factor is drawn from a distribution around 1.0 with a
standard deviation of 0.2. Panel A of Figure~\ref{sim} shows the
almost imperceivable transient component (between phase fraction 1.2
and 1.3) at its maximum. In this case, the peak of the component was
chosen to be twice the standard deviation of the noise added to the
profiles. Despite the apparently subtle nature of the profile
variation, it can be clearly seen in the corresponding variability map
(Panel B of Figure~\ref{sim}). This is because GP regression is
sensitive to even faint trends that persist over multiple data points.
\\
To mimic the case in which the profile variability has a
time-scale comparable or shorter than the observing cadence, we
shuffled the pulse profiles so that the growth and recession of the
transient component was no longer coherent. The resulting variability
map shown in Panel C of Figure~\ref{sim}. Although the magnitude of
the profile deviation is the same, the variability is not highlighted
by the detection technique. When GP regression is employed to model
the flux density variability seen in each pulse phase bin, any profile
features that occur in single observations only, will have little
effect on the model and, therefore, on the emission variability map
overall. This is desirable if the single observation has produced a
spurious pulse profile due to instrumental failure, but conversely, any
genuine profile deviations that occur in single observations may not
feature in the final emission variability map.
\section{Discussion}
\label{disc}
\subsection{New Techniques}
We have developed a new technique for measuring $\dot{\nu}$ by
employing GP regression to model the second derivative of the timing
residuals. The uncertainty of our $\dot{\nu}$ values is small because
the GP regression prediction at any point is informed by many
neighbouring data points (the extent of which is dependent on the
covariance parameters). As described in Section \ref{rotvar}, our
techniques provide fully Bayesian error estimation, which is not the case
when timing residuals are modelled using a maximum likelihood
estimator. Additionally, the method is analytical, in contrast to
previous techniques, in which $\dot{\nu}$ is numerically calculated
from the information within the span of a small window. 
\\
When fitting a non-parametric function to the timing residuals, all
but two of the nine pulsars featured in this paper were best fit by a
covariance function that contained two squared exponential kernels
plus a noise model. Table~\ref{gpparam} shows the optimised covariance
parameters when fitting the timing residuals with one and with two
kernels in the covariance function. Where the data is best fit by two
kernels, this suggests that at least two physical processes are
responsible for the timing noise that we observe.
\\
Using the above technique to measure $\dot{\nu}$, along with our
method to map pulse profile shape changes, we have been able to
reproduce the quasi-periodic variability already observed in
PSR~J0742$-$2422 \citep{2010Sci...329..408L, 2013MNRAS.432.3080K} and
PSR~J1830$-$1059 \citep{2010Sci...329..408L}.
\\
A very useful property of a GP, is that it can combine derivative and
integral observations. If, in the future, we are able to successfully use pulse profile
variability as a proxy to calculate $\dot{\nu}$ values, these data can be
easily combined with timing residuals in further GP regression.
\subsection{PSR~J1602$-$5100}
The above techniques have also uncovered a striking new example of
correlated $\dot{\nu}$ and profile shape variability in
PSR~J1602$-$5100. This pulsar exhibits a dramatic change in profile
shape over $\sim$ 600 days, with a simultaneous reduction in
$\dot{\nu}$. Such sudden and dramatic changes have previously been
attributed to exterior material entering the pulsar magnetosphere
\citep{2008ApJ...682.1152C, 2013ApJ...766....5S,
  2014ApJ...780L..31B}. The reconfiguration in current within the
magnetosphere, induced by the introduction of external material, would
simultaneously affect the braking torque and hence $\dot{\nu}$. The
value of $\dot{\nu}$ of PSR~J1602$-$5100 is observed to decrease with
the appearance of a new profile component. This is comparable to the
2005 event seen in PSR~J0738$-$4042, hypothesised to be caused by an
asteroid encounter \citep{2014ApJ...780L..31B}. The changes observed
in PSR~J0738$-$4042 have persisted, whereas PSR~J1602$-$5100 returned to
its previous state after $\sim$ 600 days. The change in $\dot{\nu}$
over this period can be approximated as a step-function, and
interpreted as a reduction in the total outflowing plasma above the
polar caps. The magnitude of the current change can be inferred from
the change in $\dot{\nu}$ following \citet{2006Sci...312..549K}. The
difference between the pre- and post-step $\dot{\nu}$ values
corresponds to a reduction in the charge density $\rho$ of $\sim 9
\times 10^{-9} $ C cm$^{-3}$, where $\rho =
3I\Delta\dot{\nu}/R_{pc}^{4}B_{0}$, the moment of inertia $I$ is taken
to be 10$^{45}$ g cm$^{2}$, the magnetic field $B_{0} = 3.2 \times
10^{19}\sqrt{-\dot{\nu}/\nu^{3}}$ G, polar cap radius $R_{pc} =
\sqrt{2\pi R^{3}\nu/c}$ and where the neutron star radius $R$ is taken
to be $10^{6}$ cm. We can relate the difference in charge density
associated with the two $\dot{\nu}$ states to mass supplied to the
pulsar, by multiplying it by the speed of light, the polar cap area
and the duration of the new spindown state. Over 600 days (the
duration of the dramatic profile and $\dot{\nu}$ changes), this
amounts to $\sim 10^{14}$~g, which lies within the range of known
solar system asteroid masses, and is consistent with the mass range of
asteroids around neutron stars proposed by
\citet{2008ApJ...682.1152C}.
\subsection{Correlated Variability}
\begin{figure*}
\begin{center}
\includegraphics[width=120mm]{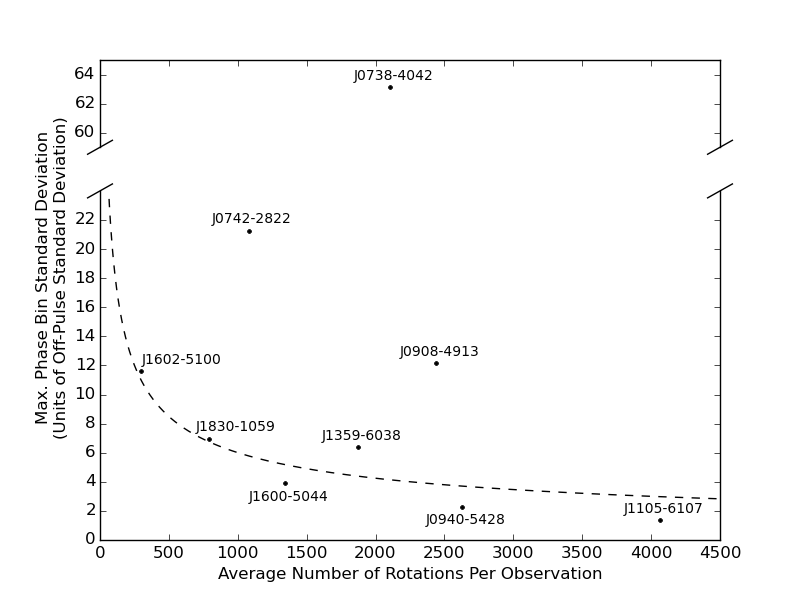}
\caption{The maximum amount of deviation of a pulsar's profile as a
  function of the average number of rotations per observation. The
  dashed line shows the $y=1/\sqrt{x}$ function that best fits all
  data points with the exception of PSR~J0738$-$4042, PSR~0742$-$2822 and PSR~J0908$-$4913.}
\label{rotvsdev}
\end{center}
\end{figure*}

The present interpretation of the correlated long-term variations
observed in emission and rotation, involves charged particle currents
in the pulsar magnetosphere \citep{2006Sci...312..549K}; distinct
pulsar states can be explained by differing levels of magnetospheric
plasma. Changing plasma levels are expected to modify both the
material outflow along open field lines at the polar cap, and the
subsequent emission produced. Plasma variations would also vary the
braking torque on the pulsar, and we would expect to see a change in
$\dot{\nu}$ accompanying any significant change in emission. The
correlated variability that has been previously observed in pulse
profiles and rotation are, therefore, evidence of intrinsic
processes. 
\\
In this paper, we have demonstrated that pulse profile shape can be
shown to correlate with $\dot{\nu}$, although the relationship does
not appear to be simple.  Of the nine pulsars analysed in this paper,
seven show significant changes in pulse profile shape to various
degrees and on various time-scales. Of those, five also show some
degree of $\dot{\nu}$ variation that could be considered
correlated. It is particularly interesting that no change in
$\dot{\nu}$ is seen to accompany the dramatic profile shape change
seen in PSR~J0738$-$4042.
\\
When interpreting these results, we must consider the possibility
that the relationship beween $\dot{\nu}$ and pulse profile variability
is more complex than anticipated and/or that the relationship is
simple, but that we are unable to accurately record the variability
involved. When analysing the rotational variations, for example, we
have made the hypothesis that timing noise is due to changes in the
braking torque on the pulsar, observable as changes in
$\dot{\nu}$. Other possible sources of timing noise are, inadequate
calibration of the raw observations
\citep[e.g.][]{2006ApJ...642.1004V}, and failure to correct for
variations in the interstellar dispersion \citep[e.g.][]{You21062007}. With
regard to the pulse profile variability analysis, the simulation
described in Section~\ref{others} demonstrates that trends that
persist over multiple observations can be detected by the techniques
described in this work, providing they have a magnitude above $\sim$
two times the level of the observation noise. Therefore, profile faint
variability that also has a short time-scale will often go
undetected. Additionally, when testing the correlation between the
variability of flux density in individual (or small groups of) pulse
phase bins and the spindown rate, we acknowledge that the latter is
measured much more accurately than the former. Longer observation
times would, in general, produce more stable pulse profiles and allow
extraordinary deviations to be more easily identified. The standard
deviation of the mean profile is proportional to 1/$\sqrt{n}$, where
$n$ is the number of pulses it includes. As each pulsar has a
different rotational period, and has observations of varying length,
the degree of their profile stability is also expected to be
different. Figure~\ref{rotvsdev} plots the average number of rotations
in an observation against the maximum amount of deviation in the pulse
profile. It is interesting to note that most of the pulsars featured
in this paper follow the expected trend, with the stark exception
being PSR~J0738$-$4042; sections of the pulse profile clearly remain
unstable since the dramatic changes undergone in 2005. PSR~J0742$-$2822
and PSR~J0908$-$4913 also show substantially more deviation than that
expected from their average observation length.
\subsection{The Relationship Between Shape and Mean Flux Density}
We have shown that changes in the shape of the pulse profile can be
accompanied by changes in its mean flux density. This is seen most
clearly in PSR~J1830$-$1059, in which the narrow emission state has a
mean flux density around 1.4 times larger than the wide state. In
other words, the flux density of the leading edge component and of the
peak are in anti-correlation. Such anti-correlation (and correlation)
is also seen in other pulsars, most notably PSR~J0738$-$4042 and
PSR~J1602$-$5100.
\\
For PSR~J1830$-$1059, it is possible to predict the spindown rate from
the radio flux density received from the pulsar. In this respect,
parallels can be drawn between the behaviour of  PSR~J1830$-$1059 and
that of the intermittent pulsars; both have two states in which higher
emission is coupled to more rapid spindown.
\\
The relationship between pulse profile shape and mean flux density
cannot be easily investigated in most pulsars, because large
variations in pulse profile flux density, due to refractive
scintillation, are ubiquitous; the pulse profile varies as a whole,
and its shape is maintained. For some pulsars, these flux density
variations are coherent over multiple observations, i.e., on a
time-scale of hundreds of days. As these variations are thought to be
primarily due to effects of propagation, we expect, and find, that
pulse profile flux density and $\dot{\nu}$ are not well correlated in
general. It is for this reason that the correlation maps produced in this work
show the relationship between the normalised pulse profiles and
$\dot{\nu}$. It should be noted that the nature of the normalisation
process itself can result in some apparent anti-correlation between
different phases of the pulse profile.
\\
When considering the changing shape of a pulse profile, it is
noteworthy that throughout the normalised flux density plots in this
paper, we see varying levels of deviation across the pulse
profile. For example, the standard deviation in one leading edge
component of PSR~J0738$-$4042 is $\sim$ 63 times larger than the mean
standard deviation seen across the off-pulse phase bins (Panels A and
B of Figure~\ref{0738_profile}). This is in contrast to other sections
of the profile that show much less deviation. Further investigation of
single pulses would elucidate the differing levels of deviation across
a pulse profile. This fact also offers the opportunity to improve
pulsar timing by preferentially employing the most stable sections of
the pulse profile when performing template matching.
\section{Conclusions}
We have analysed 168 pulsar datasets, each spanning up to eight years
in length, and presented results from nine pulsars. We have employed
GP regression as part of new analysis techniques in order to model
pulse profile variability. Simulations have shown that this is most
easily detected if the profile deviations occur over multiple
observations and have a magnitude at least twice the level of the
observational noise.
\\
GP regression was also used to infer $\dot{\nu}$ under the assumption
that all timing noise is the result of unmodelled changes in
$\dot{\nu}$. Our variability detection techniques have accurately reproduced
known pulsar variability, and also discovered some clear new examples. The most
notable new variability was found in PSR~J1602$-$5100; dramatic
pulse profile changes along with a $\sim$ 5\% rise and fall in $\dot{\nu}$ occur
simultaneously over a $\sim$ 600 day span.
\\
The correlation between $\dot{\nu}$ and changes in pulse profile shape
is clear in some pulsars, but not in others. We must consider that one
or more of the following is true: (i) The intrinsic relationship
between $\dot{\nu}$ and pulse profile variability is possibly more
complex than has been postulated previously. (ii) Our ability to
detect pulse profile variability is often insufficient to show the
underlying correlation with $\dot{\nu}$. (iii) Our hypothesis that
unmodelled changes in $\dot{\nu}$ are primarily responsible for
timing noise, may be invalid. Problems regarding the detection of
pulse profile variability will be mitigated by more sensitive
instruments and longer integration times.
\\
Finally, we have observed a strong relationship between the shape
changes of PSR~J1830$-$1059 and its mean flux density, both of which are
correlated with $\dot{\nu}$. One pulse profile shape has a mean flux
density around 1.4 times that of the other, meaning that the
monitoring of either the profile shape or the mean flux density would
permit the removal of most of the timing noise in this pulsar.
\section*{Acknowledgments}
The Parkes radio telescope is part of the Australia Telescope National
Facility which is funded by the Commonwealth of Australia for
operation as a National Facility managed by CSIRO. Data taken at Parkes
is available via a public archive. P.R.B. is grateful to
the Science and Technology Facilities Council, the Royal Astronomical
Society and CSIRO Astronomy and
Space Science for support throughout this work. The authors would like
to thank the anonymous reviewer for their valuable comments and
suggestions to improve the quality of the paper. \\






\bsp	
\label{lastpage}
\end{document}